\documentclass[amsmath,amssymb,aps,pra,twocolumn,10pt]{revtex4-2}

\usepackage{graphicx}
\usepackage{hyperref}
\usepackage{xcolor}

\begin{document}

\title{Space-time singularities in spatially-chirped Laguerre-Gaussian beams of any order}

\author{Spencer W. Jolly}
\affiliation{Service OPERA-Photonique, Université libre de Bruxelles (ULB), Brussels, Belgium}
\email{spencer.jolly@ulb.be}
\date{\today}

\begin{abstract}
The electric field distributions and space-time singularity curves are computed for ultrashort pulsed Laguerre-Gaussian laser beams having spatial chirp. Due to the breaking of cylindrical symmetry by the spatial chirp, the singularities trace complicated curves in space-time, which also vary for different combinations of radial and vortical orders. Analytical solutions are mostly presented along with a recipe for numerically calculating higher orders. The behavior of the singularities upon propagation is also shown, along with a discussion of the extension towards few-cycle pulses. These results are an example of how a simple physical scenarios can result in highly complicated singular behavior in space-time.
\end{abstract}

\maketitle

\section{introduction}

Space-time couplings in ultrashort laser pulses~\cite{akturk05,akturk10} are becoming a more common and interesting lever for not only controlling and structuring the light beams themselves~\cite{shenY23,liuX24}, but also for controlling light-matter interaction. The level of sophistication of space-time pulse shaping is rapidly increasing~\cite{mounaix20}, allowing for high-level design and high-order structuring. As an example, space-time optical vortices (STOVs)~\cite{sukhorukov05,jhajj16,wanC23,porras23-1,bekshaev2024spatiotemporal} are a generalization of traditional optical vortices in the space-time domain, they contain space-time singularities and transverse angular momentum~\cite{chong20}. Going beyond the simplest case, STOVs have been arbitrarily oriented~\cite{wangH21,zangY22,liuX24-2} and structured~\cite{chenW22,caoQ24} in more complex ways. The downside of course, of these very advanced and powerful techniques, is that they are generally complicated, involve significant losses, may have low fidelity, and require components with low damage thresholds.

Spatial chirp (SC)~\cite{gu04}, where the component colors are separated along one transverse coordinate, is one of the simplest and most common space-time couplings generated by simple optical components. It has been taken advantage of most notably for the attosecond lighthouse technique~\cite{vincenti12,wheeler12,kim13-2,quere14,auguste16}. SC has recently been shown to produce space-time vortex structures in a number of configurations where analytical solutions could be found~\cite{porras23-3,jolly24-1,porras25-2}. The hallmark of these recent works is that they were combinations of paraxial propagation modes of free space that are in some way higher-order than the standard Gaussian beam mode, i.e. the Hermite-Gaussian beam~\cite{porras25-2}, the optical vortex~\cite{porras23-3}, and a radially-polarized beam~\cite{jolly24-1}. In this context, the theoretical work is emulating extremely simple experimental scenarios whereby purely spatial shaping is combined with the generation of spatial chirp---possible with either lenses or prisms to apply angular dispersion (and pulse-front tilt, PFT)~\cite{bor93,pretzler00,torres10}, and subsequent propagation or focusing to convert to spatial chirp. All cases resulted in different modifications to the existing optical singularities.

\begin{figure}[htb]
\centering
\includegraphics[width=86mm]{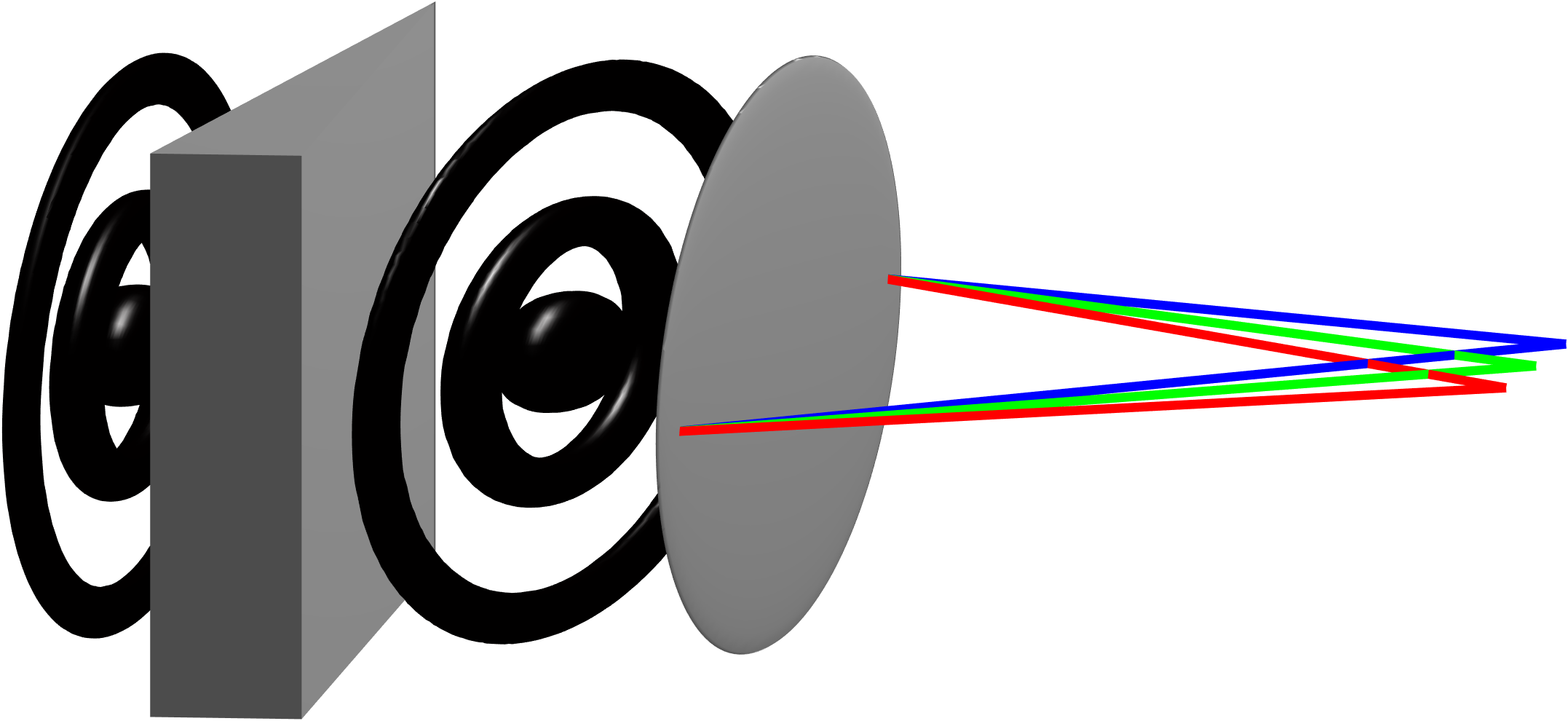}
\caption{Sketch of the physical scenario under consideration. An ultrashort Laguerre-Gauss beam (pictured here with $\{l,p\}=\{0,2\}$) is sent through a dispersive element (here a prism) to acquire pulse-front tilt (and angular dispersion). When focused, such a pulse-beam will have spatial chirp. We study in this work how the field behaves near such a focus, i.e. an ultrashort Laguerre-Gaussian beam with spatial chirp.} 
\label{fig:1}
\end{figure}

In this work we consider a more general case of Laguerre-Gaussian (LG) beams of any order combined with spatial chirp. That specific scenario is shown in more detail in Figure~\ref{fig:1}. This case is interesting because, similar to the cases of the optical vortex or radial polarization, the spatial chirp breaks the cylindrical symmetry of the LG mode. However, now there are a larger number of optical singularities and potentially singularities of different types. It will be seen that this has a more exotic effect on the singularities, pushing them to have a space-time character but also uncommon and discontinuous behavior. After introducing the relevant optical physics, we will combine analytical and numerical methods to describe a few examples. This work adds an important understanding of how space-time singularities behave and can be created or controlled. We note that our case is in free-space (i.e. no medium~\cite{hyde23}), has no additional chirp~\cite{novikov25}, and doesn't involve frequency conversion~\cite{gaoX25}, emphasizing the ability and potential of simple techniques, and enriching the context of such simple techniques together with the more complex and arbitrary techniques mentioned previously.

\section{theoretical tools}

The most compact solution of the paraxial wave equation for the pulsed Gaussian beam of symmetric beam waist $w_0$ and pulse duration $t_0$ (frequency bandwidth $\Omega_0=2/t_0$) is, in it's frequency ($\tilde{E}$) form is

\begin{align}
\tilde{E}=e^{i\omega z/c} f e^{-\omega f\rho^2/\omega_0} e^{-\Omega^2/\Omega_0^2}\label{eq:Gauss_freq},
\end{align}

\noindent where $\Omega=\omega-\omega_0$, $f=i/(i-\zeta)$, $\zeta=z/z_R$, and $\rho=\sqrt{x^2+y^2}/w_0$ allow for the most compact description. $w_0$ is the beam waist and $z_R=\pi w_0^2/\lambda_0=\omega_0 w_0^2/2c$ is the Rayleigh range.

The fields of an ultrashort Laguerre-Gaussian (LG) can be written in the compact notation of this article as

\begin{align}
\begin{split}
\tilde{E}^\textrm{LG}=&e^{i\omega z/c} \left(\frac{f}{|f|}\right)^{2p} f^{|l|+1} e^{-\omega f\rho^2/\omega_0} e^{-\Omega^2/\Omega_0^2}\\
&\times\left(\xi\pm i\eta\right)^{|l|}L_p^{|l|}(2|f|^2\rho^2)\label{eq:LG_freq},
\end{split}
\end{align}

\noindent where the topological charge of the LG beam is $l$ with the handedness described by the $\pm$ operator, and the radial number is $p$. The normalized Cartesian coordinates are $\xi=x/w_0$ and $\eta=y/w_0$, which are necessary to separate from $\rho$ to properly describe the vortex.

We model spatial chirp in the fields by doing the coordinate transformation $x\rightarrow x-b\Omega$ in $\tilde{E}$ and performing the Fourier-transform. For simplicity we write $B=b\Omega_0/w_0$ and we note $\omega=\Omega+\omega_0$, and regroup the terms such that all those depending on $\Omega$ are together. The spatially-chirped LG beam, still in frequency space, is

\begin{align}
\begin{split}
&\tilde{E}^\textrm{LG-SC}=e^{i\omega_0 z/c} \left(\frac{f}{|f|}\right)^{2p+|l|} f e^{-f\rho^2}\\
&\times e^{i\Omega z/c}e^{-\Omega f\rho^2/\omega_0}e^{-f\left(1+\frac{\Omega}{\omega_0}\right)\left(\frac{B^2\Omega^2}{\Omega_0^2}-\frac{2B\Omega\xi}{\Omega_0}\right)}e^{-\Omega^2/\Omega_0^2}\\
&\times \left(|f|\left(\xi-\frac{B\Omega}{\Omega_0}\right)\pm i|f|\eta\right)^{|l|}\\
&\times L_p\left(2|f|^2\left(\left(\xi-\frac{B\Omega}{\Omega_0}\right)^2+\eta^2\right)\right)\label{eq:LG_SC_freq}.
\end{split}
\end{align}

We now need to use the Fourier transform

\begin{equation}
E=\frac{e^{-i\omega_0 t}}{2\pi}\int\limits_{-\infty}^\infty d\Omega\, \tilde{E} e^{-i\Omega t} \label{eq:Fourier_transform}
\end{equation}

\noindent to calculate the fields in time. We ignore one term $\propto\Omega^3$ and another term $\propto\Omega^2\xi$ in the exponential of Eq.~\ref{eq:LG_SC_freq} that are negligible except for few-cycle pulses:

\begin{align}
\begin{split}
&E^\textrm{LG-SC}=f\left(\frac{f}{|f|}\right)^{2p+|l|} e^{-i\omega_0 t^{\prime\prime}}\\
&\times \int\limits_{-\infty}^\infty d\Omega\,e^{\Omega\left(\frac{2fB\xi}{\Omega_0} -it^{\prime\prime}\right)}e^{-\Omega^2(1+fB^2)/\Omega_0^2}\\
&\times \left(|f|\left(\xi-\frac{B\Omega}{\Omega_0}\right)\pm i|f|\eta\right)^{|l|}\\
&\times L_p\left(2|f|^2\left(\left(\xi-\frac{B\Omega}{\Omega_0}\right)^2+\eta^2\right)\right)\label{eq:LG_SC_integral},
\end{split}
\end{align}

\noindent with $t^\prime=t-z/c$ the local time along propagation, and $t^{\prime\prime}=t^\prime - if\rho^2/\omega_0$ the local time including the delay accrued due to curvature on propagation. Note of course that if $B=0$, i.e. no spatial chirp, then the Laguerre polynomial and vortex polynomial (that to the power of $|l|$) have no frequency dependence and do not contribute to the then purely Gaussian integral.

Finally, starting from Eq.~\ref{eq:LG_SC_integral} we make the following variable substitutions

\begin{align}
\tau&=\frac{\Omega_0t^{\prime\prime}}{2},\\
u&=\frac{\sqrt{1+fB^2}}{\Omega_0}\left(\frac{\xi\Omega_0}{B}-\Omega\right),\\
v&=\frac{(\xi+iB\tau)}{B\sqrt{1+fB^2}},\\
\alpha&=\frac{\sqrt{2}|f|B}{\sqrt{1+fB^2}},\\
c&=\sqrt{2}|f|\eta,
\end{align}

\noindent and complete the square in the exponent, such that the spatially-chirped LG fields in time can be written in terms of a basic space-time field and a single relatively compact integral:

\begin{align}
\begin{split}
E^\textrm{LG-SC}=&f\left(\frac{f}{|f|}\right)^{2p+|l|} \frac{e^{-i\omega_0 t^{\prime\prime}}}{\sqrt{1+fB^2}}\exp{\left[-\frac{(\tau+if\xi B)^2}{(1+fB^2)}\right]}\\
\times \int\limits_{-\infty}^\infty du&\,e^{-(u-v)^2} \left(\alpha u\pm ic\right)^{|l|}L_p^{|l|}\left(\left(\alpha u\right)^2+c^2\right),\\
=&\left(\frac{f}{|f|}\right)^{2p+|l|}\psi^\textrm{SC}\mathcal{I}_{l,p}.\label{eq:LG_SC_simplified}
\end{split}
\end{align}

\noindent $\mathcal{I}_{l,p}$ represents the integral and $\psi^{SC}$ is a universal profile for spatially-chirped fields that describes their inherent wavefront rotation (and increased size and duration, and decreased peak field). Note that the integral for the case of a Gaussian is only a constant factor $\mathcal{I}_{0,0}=\sqrt{\pi}$. Therefore the Gouy phase term and the integral contain all particularities of the given LG beam of any order.

Finding a solution $\mathcal{I}_{l,p}(v,\alpha,c)=\mathcal{I}_{l,p}(x,y,t)$ to the Gaussian integral involving the Laguerre polynomial has so far not been possible when considering any arbitrary orders. If looking in 3D, the term $c$ ($\propto y$) adds significant complexity since a sum in the Laguerre polynomial argument requires a sum of multiple polynomial pairs of each separate argument. Even in the $x-t$ plane only, i.e. $y=\eta=c=0$, and with no vorticity, i.e. $l=0$, we have not found a solution to the integral. The only analytical solution found in the past is for the pure optical vortex, with $p=0$, described in Ref.~\cite{porras23-3}. This means that with finite spatial chirp we must work with a few solutions to the lowest order scenarios (and therefore simplest) and numerical solutions to the integral in Eq.~\ref{eq:LG_SC_simplified} for higher-order cases.

\section{Analysis of the electric field profiles and their singularities}

The case of $p=0$ was studied in our past work~\cite{porras23-3}, where in the notation of this article

\begin{equation}
\frac{\mathcal{I}_{l,0}}{\sqrt{\pi}}=\left(\frac{\alpha}{2i}\right)^{|l|}H_{|l|}\left(iv\mp\frac{c}{\alpha}\right).
\end{equation}

\noindent In that case the spatial chirp caused the purely longitudinal line singularity of topological charge (TC) $l$ to separate along the $y$-axis into $|l|$ line singularities of unitary TC, now oriented both longitudinally and transversely. This can be seen in the argument of the Hermite polynomial $H_{|l|}$, which can only result in a singularity when it is purely real, i.e. when $\textrm{Re}\{v\}\propto x=0$, and those zeros are in the $y-t$ plane and are related to the zeros of the Hermite polynomial. We therefore showed that the spatial chirp could tune the orientation of the singularity (and likely the angular momentum direction), which also precessed upon propagation. Following up on that work, in this section we will consider other cases, i.e. where $p\neq0$.

\subsection{Purely radial beams}

The purely radial LG beam is where $l=0$ and $p\neq0$. The simplest case of this is of course where $p=1$. In that case we can solve the integral and find that

\begin{equation}
\frac{\mathcal{I}_{0,1}}{\sqrt{\pi}}=1-c^2-\frac{\alpha^2}{2}-\alpha^2 v^2.
\end{equation}

\begin{figure}[htb]
\centering
\includegraphics[width=86mm]{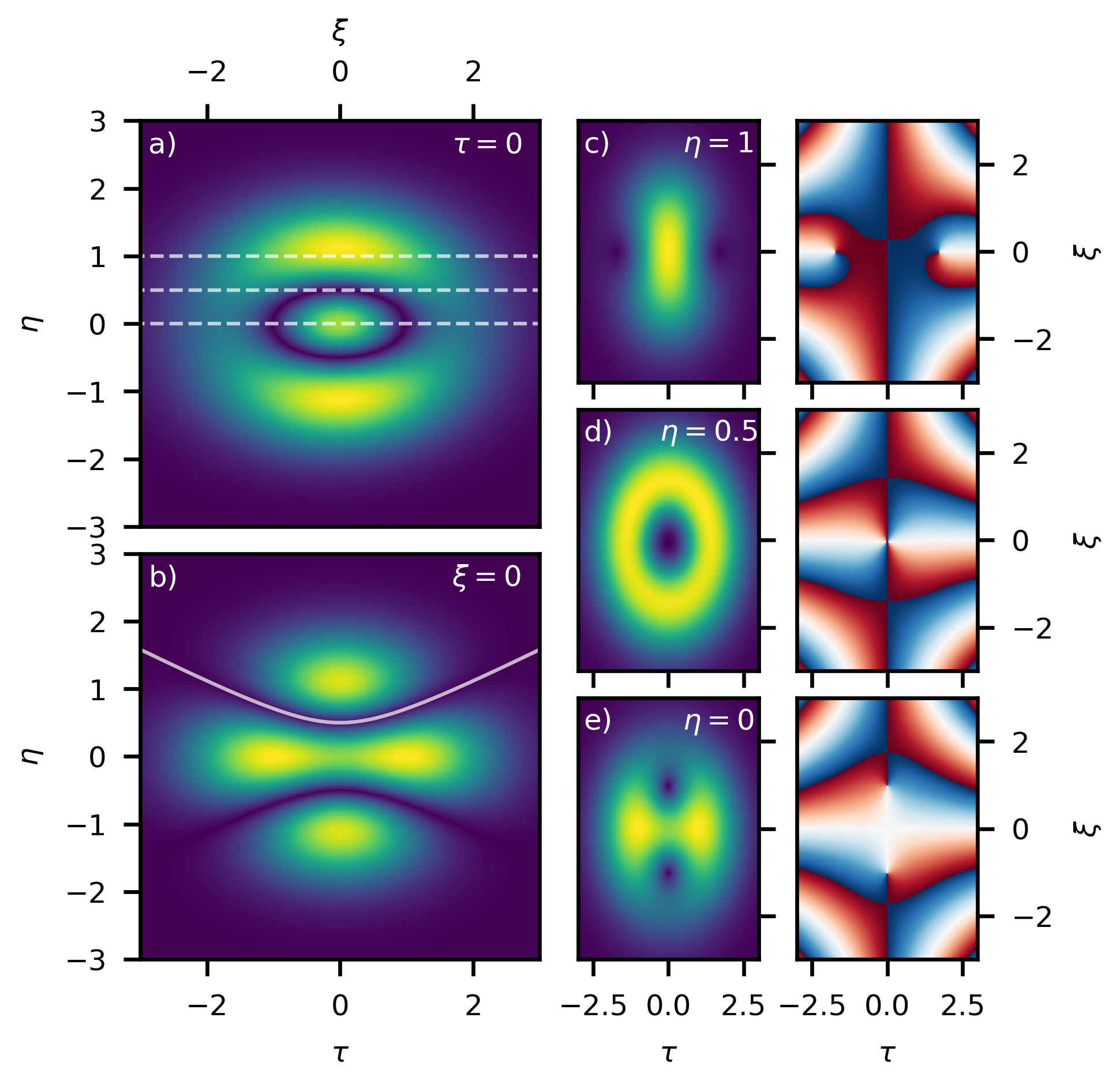}\\
\includegraphics[width=43mm]{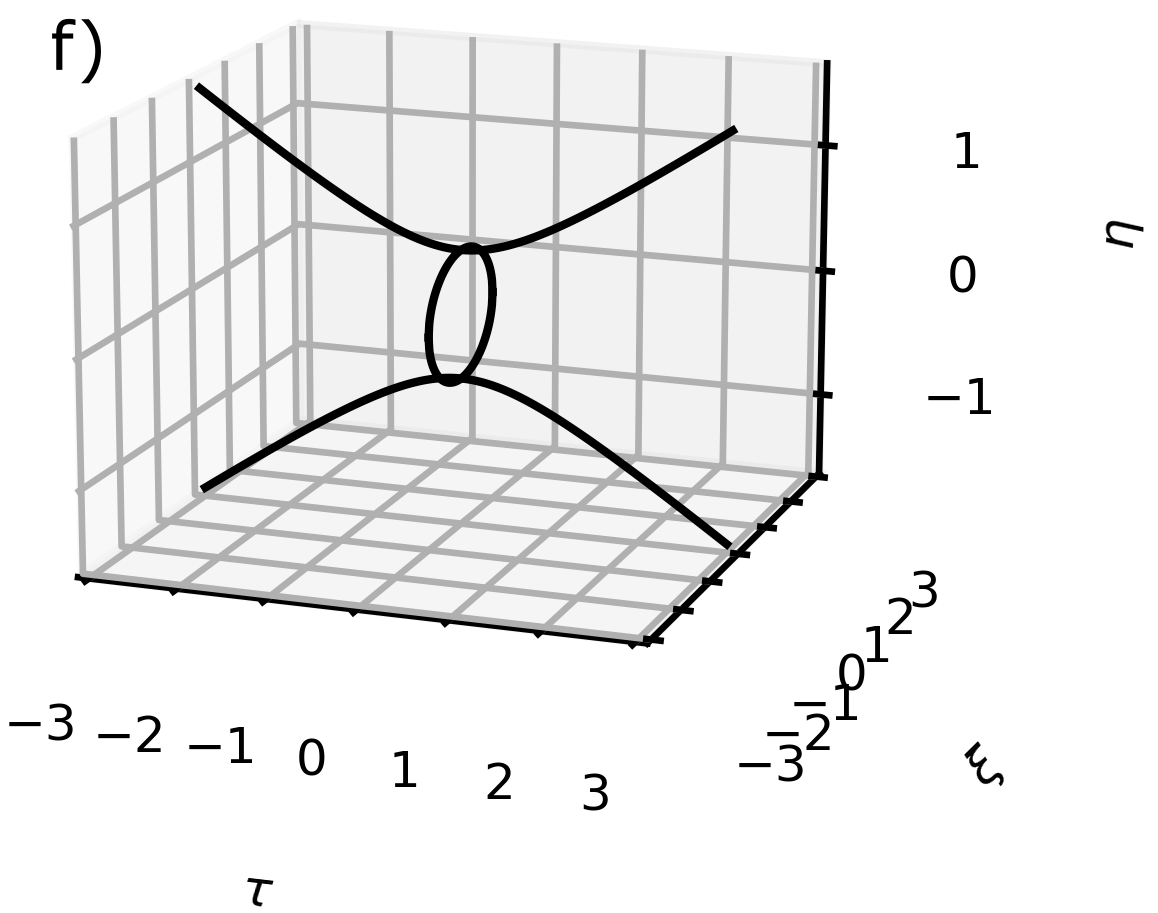}
\caption{Amplitude and phase profiles of the spatially-chirped LG beam with $l=0$ and $p=1$, at the focus ($z=0$) and with $B=1$. The amplitude is shown for cuts at $t=0$ (a) and $x=0$ (b), where the hyperbola shape is traced in a solid white line. Amplitude and phase are shown for different $y$ positions (c--e) denoted by the dashed white lines in (a), where the phase is shown on the right in each case. These cuts paint a picture of the 3D singularity curve, which is shown in (f). All amplitude profile plots are normalized to their own maximum, and all phase plots span $[\pi,\pi]$, which will be true for all figures.} 
\label{fig:2}
\end{figure}

\noindent The resultant electric field of the pulse-beam in three dimensions can be seen from a number of different perspectives in Figure~\ref{fig:2} where of course the singularities and their space-time vortex nature are the most interesting and novel element.

The $2p=2$ singularities in this space-time coupled structured light beam, where the intensity is also zero, are at positions in space-time that satisfy the following equation:

\begin{equation}
\frac{(\xi+iB\tau)^2}{(1+fB^2)^2}+\eta^2=\frac{1}{2|f|^2}-\frac{B^2}{2(1+fB^2)},
\end{equation}

\noindent which is the result of $\mathcal{I}_{0,1}=0$. The most intuitive position is at $z=0$ where $f=1$ and is therefore purely real. In that further simplified case, the singularities are at 

\begin{equation}
\frac{(\xi+iB\tau)^2}{(1+B^2)^2}+\eta^2=\frac{1}{2(1+B^2)}.\label{eq:singularities}
\end{equation}

\noindent Although this was a trivial step from the previous equation, this now allows to easily see how the singularities come about. Note that there are no solutions to the above equation where both $\xi$ and $\tau$ are different from zero, since in that case there would be an imaginary term with no pair to cancel it out. This hints at the singularities being in planes.

At $t=0$, the singularities trace an ellipse in $x$ and $y$ where the ellipticity is determined by $B$ (Fig.~\ref{fig:2}a). With $x=0$ the singularities form hyperbolae in $y$ and $t$ where the eccentricity is determined by $B$ (Fig.~\ref{fig:2}b). It happens that the ellipse and the hyperbolae overlap exactly at the points $\{\xi,\eta,\tau\}=\{0,\pm1/\sqrt{2(1+B^2)},0\}$, which is seen at the points $\eta=\pm0.5$ in Fig.~\ref{fig:2}. This can be thought of as a discontinuous break, where for all positive and negative times the singularities trace the hyperbolae, but at the instant of $t=0$ they are opened onto the ellipse. Viewed in the $x-t$ plane for different $y$ values (Fig.~\ref{fig:2}c--e) the singularities are dispersed only along $x$ or only along $t$ depending on if the plane intersects the ellipse or a hyperbola. Figure~\ref{fig:2}f shows the line singularities traced in three dimensions, which we know exactly from Eq.~\ref{eq:singularities}.

Beyond the position of the singularities, it is important to note their space-time vortex character. Along the lines traced in Fig.~\ref{fig:2}f there is always a space-time vortex. In almost all of time and space the TC of each vortex is $\pm 1$, except for the two points where the ellipse and hyperbolae intersect, where the TC is $\pm 2$. This can be easily noticed in comparing the two single-charged vortices in Fig.~\ref{fig:2}c and e compared to the doubly-charged vortex in Fig.~\ref{fig:2}d, the latter of which is at the $y$ position of the intersection. The magnitude of $B$ does not change the topology of the singularities for $p=1$, and only scales them in some fashion. We saw that with spatially-chirped Hermite-Gaussian beams~\cite{porras25-2} changing the spatial chirp could reorganize the singularities, but we do not see that here so far.

Predictably, the situation becomes more complicated for larger $p>1$. As mentioned, we do not have a general analytical solution for any $p$, but we can perform the integral manually for the next case of $p=2$. We find

\begin{align}
\begin{split}
\frac{\mathcal{I}_{0,2}}{\sqrt{\pi}}=&1-2c^2+\frac{c^4}{2}+\frac{\alpha^2}{2}(c^2-2) + \frac{3\alpha^4}{8}\\
&+\left(\alpha^2(c^2-2)+\frac{3\alpha^4}{2}\right)v^2+\frac{\alpha^4}{2}v^4.
\end{split}
\end{align}

\begin{figure}[htb]
\centering
\includegraphics[width=86mm]{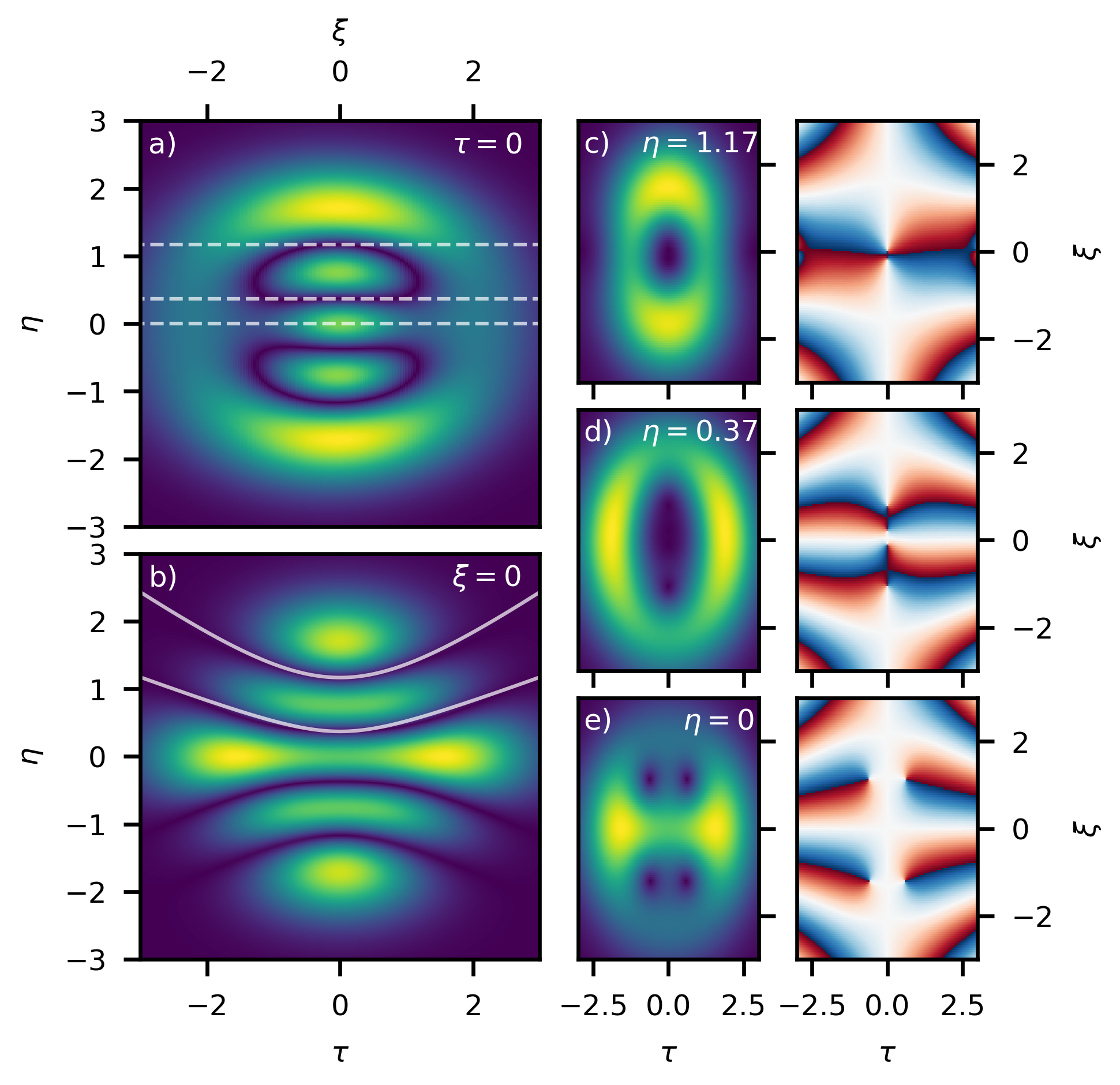}
\caption{Amplitude and phase profiles of the spatially-chirped LG beam with $l=0$ and $p=2$, at the focus ($z=0$) and with $B=1$. The amplitude is shown for cuts at $t=0$ (a) and $x=0$ (b), where the two hyperbola-like curves are traced in solid white lines. Amplitude and phase are shown for different $y$ positions (c--e) denoted by the dashed white lines in (a), where the phase is shown on the right in each case.} 
\label{fig:3}
\end{figure}

\noindent The field for this case at $z=0$ ($f=1$) and for $B=1$ is shown in Fig.~\ref{fig:3} at a number of time slices. We see similarities in that there are hyperbolae and ellipses. But, as seen in the complexity of $\mathcal{I}_{0,2}$ the location of these singularity curves is not as intuitive.

If both $\xi$ and $\tau$ are zero, i.e. $v=0$, then the singularities are located at $2p=4$ positions along $y$ according to the zeros of the equation, depending on constants and $\alpha$. For the case of $z=0$ ($f=1$) and $B=1$, then $\alpha=1$ as well, and the singularities are located at $\{\xi,\eta,\tau\}=\{0,\pm\sqrt{3/4\pm\sqrt{3/8}},0\}$. This is seen in the positions at $\eta\approx\pm0.371$ and $\pm1.167$ in Fig.~\ref{fig:3}. The more interesting aspect at $t=0$, in Fig.~\ref{fig:3}a, is that there are two ellipse-like singular curves that are centered above and below the $x-t$ plane. These are deformed ellipses, which are a solution of a fourth-order equation that has mixed terms of $\xi$ and $\eta$.

Again regarding the space-time vortices, there is once again an intersection between the hyperbola-like curves (Fig.~\ref{fig:3}b) and the ellipse-like curves such that there is a doubly-charged space-time vortex, seen in both Fig.~\ref{fig:3}c and d. We also see evidence of the fourth-order ellipse curve in Fig.~\ref{fig:3}d, since we also see singularities at $\pm\xi$ positions, since the curve briefly dips below $\eta=0.37$ and then rises again. In the $p=1$ case where it was a perfect ellipse, in Fig.~\ref{fig:2}d, we did not see this.

Going away from $t=0$ (or $\xi=0$), but still with $z=0$ ($f=1$), we find that at some positive and negative times there is another single ellipse-like curves located roughly in the $x-y$ plane, centered at $x=y=0$. This is seen in the four singularities in Fig.~\ref{fig:3}e which are the intersections of those two ellipses in that plane. Considering that there are also now $2p=4$ hyperbolae, we can see that at $\tau=0$ the two ellipse-like curves connect the outer two hyperbolae, and at the positive and negative times the single ellipses connect the inner two hyperbolae. At those intersections the topological charge of the space-time singularities is $2$, where otherwise the TC is always $1$. We will return to this behavior soon.

\begin{figure}[htb]
\centering
\includegraphics[width=86mm]{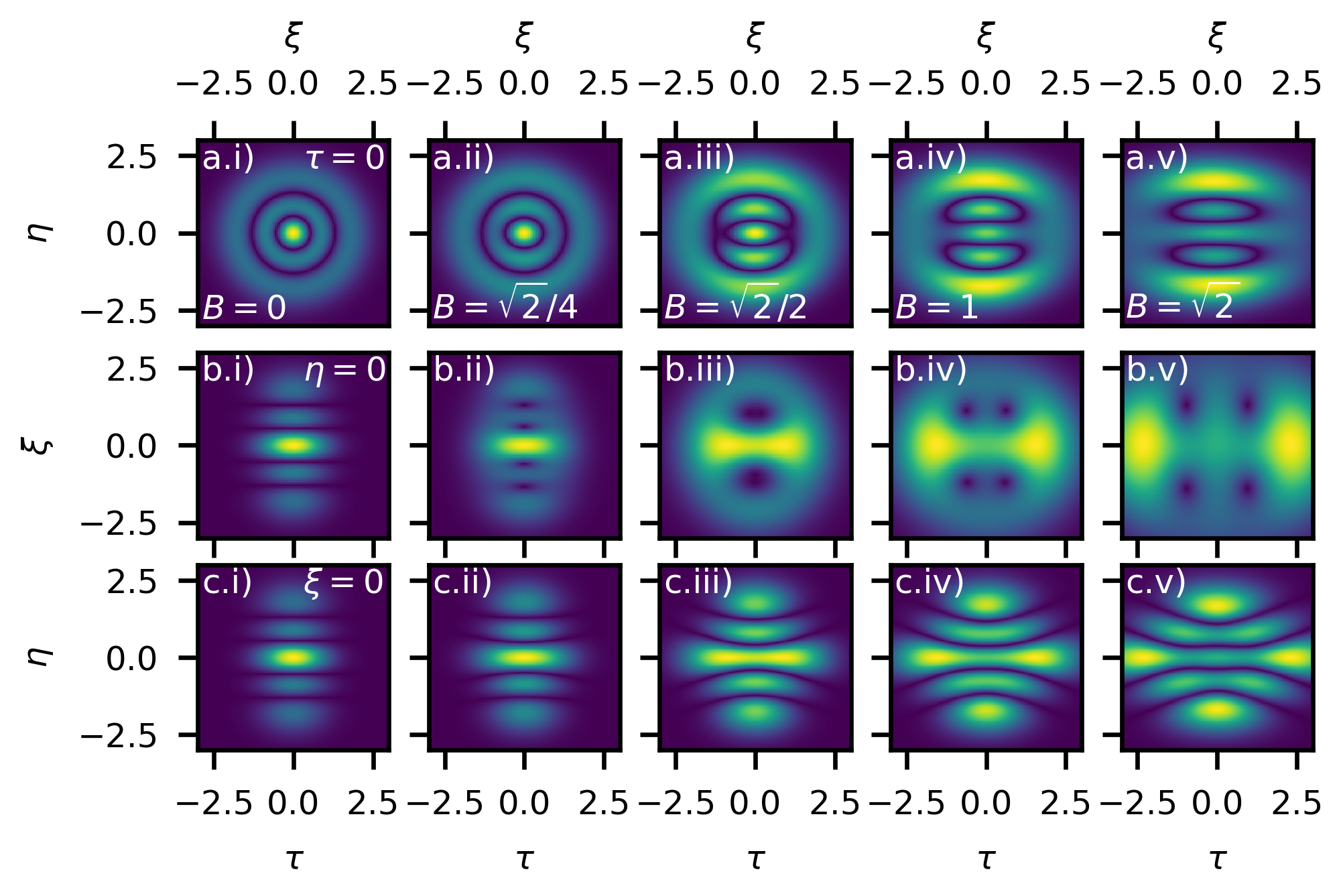}\\
\includegraphics[width=43mm]{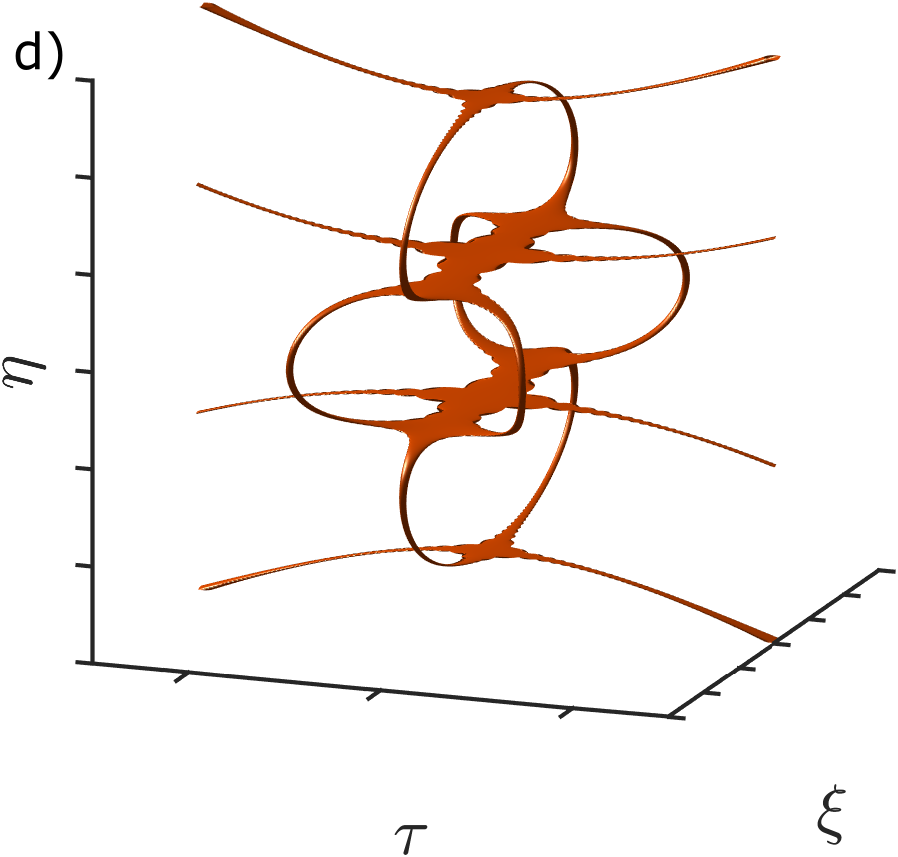}%
\includegraphics[width=43mm]{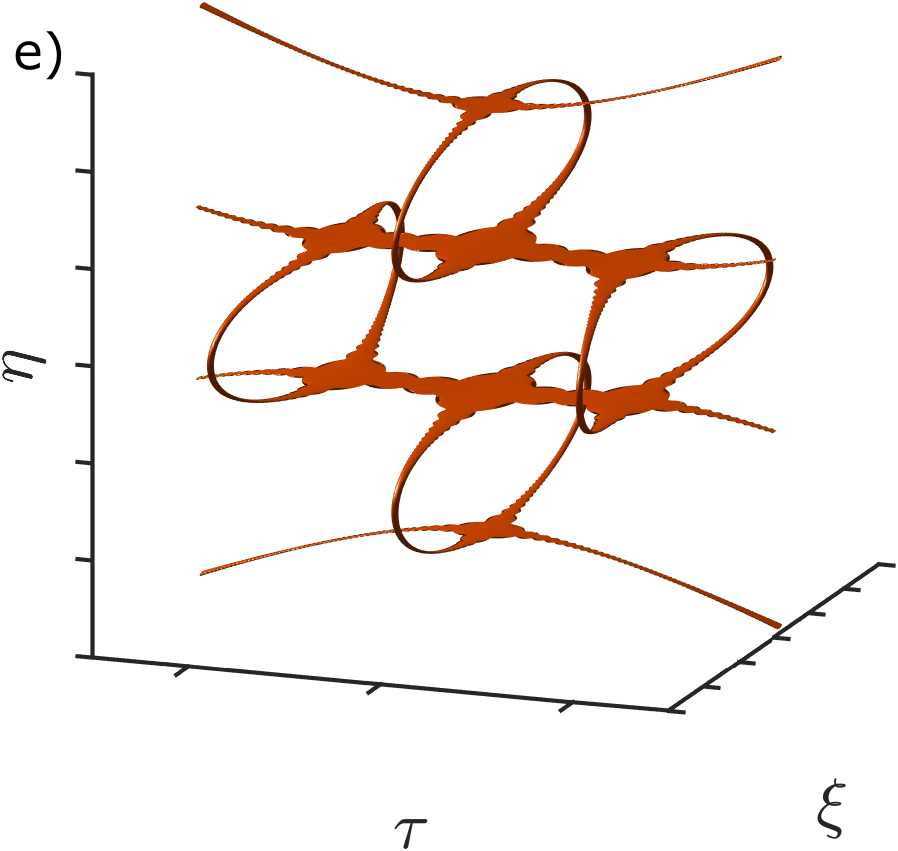}
\caption{Amplitude profiles of the spatially-chirped LG beam with $l=0$ and $p=2$, at the focus ($z=0$). The three plane cuts (a--c) are shown for a number of $B$ values (i--v). Isosurfaces at $1/5000$ of the peak intensity are shown for $B=1$ (d) and $B=\sqrt{2}$ (e) tracing the singularities where they leave the cardinal planes.} 
\label{fig:4}
\end{figure}

Contrary to the case of $p=1$ visualized in Fig.~\ref{fig:2}, where we did not see a topological difference in changing the spatial chirp $B$, with $p=2$ we do see a change. Figure~\ref{fig:3} was with $B=1$, which was an important number in previous work~\cite{porras23-3,porras25-2}. Now in Fig.~\ref{fig:4} we show, still at $z=0$ ($f=1$), the three plane slices of 3D space (a--c) for five values of spatial chirp $B=\{0,\sqrt{2}/4,\sqrt{2}/2,1,\sqrt{2}\}$ (i--v). It turns out in this case that $B_0=\sqrt{2}/2$ is the important number in this case, whereby singularities combine and above and below the special value a transition occurs in the arrangement of the singularities.

Revisiting the case without spatial chirp, the singularities are along closed circular planes with their axis centered on $\{x,y\}=\{0,0\}$ as in Fig.~\ref{fig:4}a.i, which appear as lines in both Fig.~\ref{fig:4}b.i and c.i. Increasing the spatial chirp, but staying below $B_0$, the singularities begin to close up in the $x-t$ plane (Fig.~\ref{fig:4}b.ii) but stay at $t=0$ and distributed along $x$, and open up in the $y-t$ plane (Fig.~\ref{fig:4}c.ii). At $B_0$, the singularities overlap in the $x-t$ plane (Fig.~\ref{fig:4}b.iii) forming two doubly-charged space-time vortices. Above $B_0$ the singularities in the $x-t$ plane (Fig.~\ref{fig:4}b.iv--v) are distributed along $t$, reminiscent of the Hermite-Gaussian case~\cite{porras25-2}. Interestingly, the circular curves at $t=0$ with increasing spatial chirp (Fig.~\ref{fig:4}a.ii--v) deform horizontally into ellipses but also displace vertically, whereby above $B_0$ they no longer touch. Around $B_0$ the ellipses seem the most distorted (compare Fig.~\ref{fig:4}a.iii--iv to a.ii and a.v), but on the contrary as $B$ is increased the hyperbolae in the $y-t$ plane become more distorted (Fig.~\ref{fig:4}c.v). These dynamics can be mathematically traced to the fourth-order nature of the integral and therefore the singularity curves in space-time.

With a spatial chirp $B\leq\sqrt{2}/2$ the singularity curves stay in the cardinal planes, and are fully captured in Fig.~\ref{fig:4}i--iii. However, above that value, the singularities leave the cardinal planes and trace more complicated trajectories such that they are not captured in the plane cuts in Fig.~\ref{fig:4}iv--v. We show isosurfaces for those two spatial chirp values in Fig.~\ref{fig:4}d--e showing that complicated behavior, where in fact not only are the curves not in cardinal planes, but they are not contained in planes at all and trace 3D curves. We could not draw this precisely as for Fig.~\ref{fig:2}f since we haven't solved specifically for the curves and therefore must use the isosurfaces.

\subsection{Vortical and radial beams}

In the simplest case where both $l,p\neq0$, i.e. $l=\pm1$ and $p=1$ we can manually calculate the integral to be

\begin{align}
\begin{split}
\frac{\mathcal{I}_{\pm1,1}}{\sqrt{\pi}}=&\pm ic\left(2-c^2+\frac{\alpha^2}{2}\right)+\alpha\left(2-c^2-\frac{3\alpha^2}{2}\right)v\\
&\mp ic\alpha^2v^2-\alpha^3v^3.
\end{split}
\end{align}

\begin{figure}[htb]
\centering
\includegraphics[width=86mm]{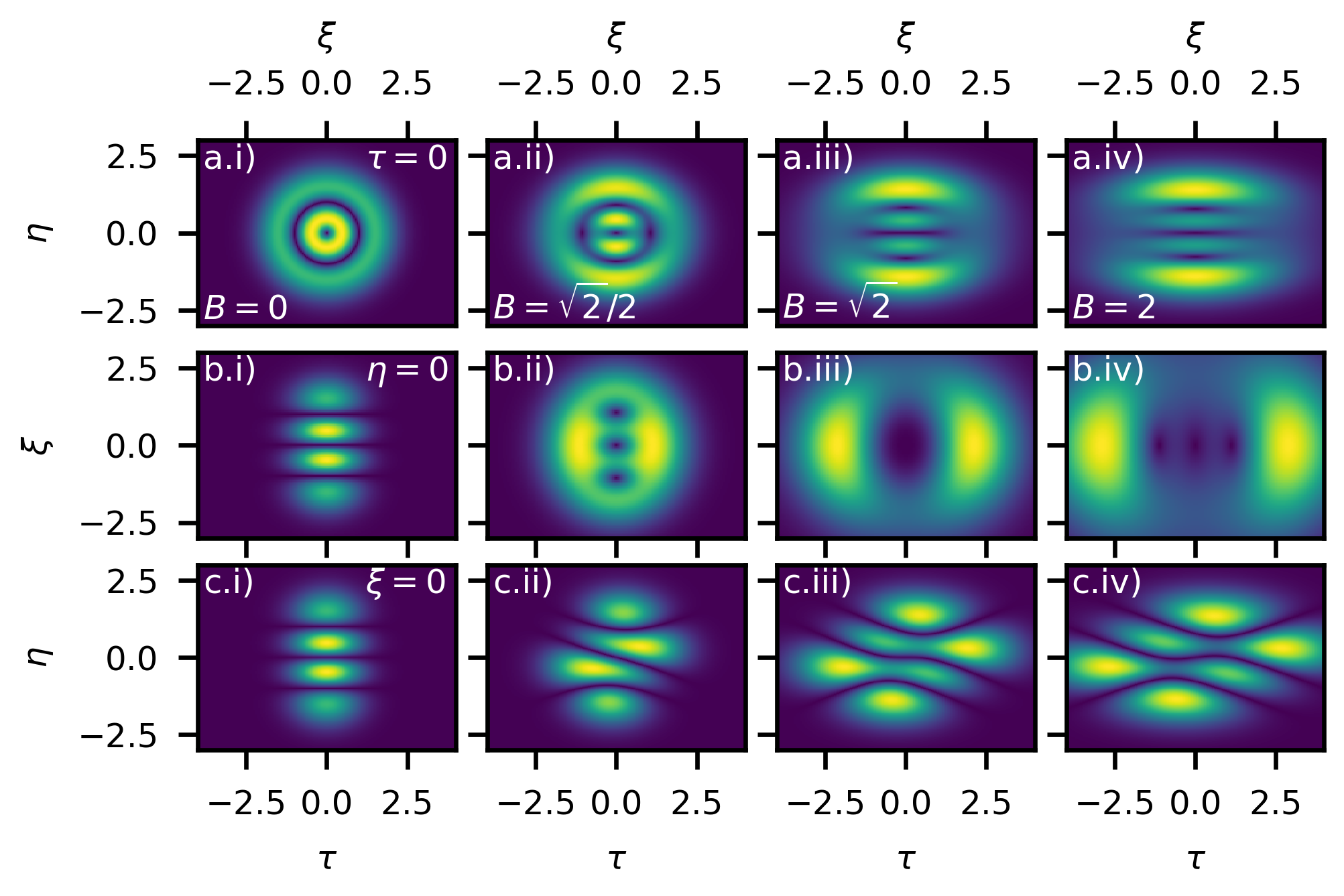}
\caption{Amplitude profiles of the spatially-chirped LG beam with $l=1$ and $p=1$, at the focus ($z=0$). The three cuts (a--c) are shown for a number of $B$ values (i--iv).} 
\label{fig:5}
\end{figure}

\noindent There are a number of new properties of the integral, which reflect upon the singularities and electric field profile. These are summarized first in Fig.~\ref{fig:5} at $z=0$ ($f=1$), where the three plane slices of 3D space (a--c) are shown for four values of spatial chirp $B=\{0,\sqrt{2}/2,\sqrt{2},2\}$ (i--iv). There are now odd powers of $v$ and $c$ in the integral solution, such that $|l|$ contributes singularly to the number of singularities while $p$ contributes doubly, so there are 3 singularities in play. We also note that a line singularity must pass through the origin $\{x,y,t\}=\{0,0,0\}$, which was not true with purely radial beams. This is intuitive, since the vortical beam without spatial chirp has a purely longitudinal vortex that also goes through the origin~\cite{porras23-3}. At $\{x,t\}=\{0,0\}$ two singularities are at $y=\pm\sqrt{1-B^2/2(1+B^2)}$ ($\pm\sqrt{3/4}\approx\pm0.866$ for $B=1$) in addition to the one at $y=0$ (see Fig.~\ref{fig:5}a).

The singularities at $y=c=0$ are at different points in the $x-t$ plane depending on the magnitude of $B$ (see Fig.~\ref{fig:5}b). If $y=0$ and $z=0$ ($f=1$), then this transition occurs at $B_0=\sqrt{2}$, below which the singularities are purely along $x$ at $t=0$, and above which they are purely along $t$ at $x=0$. This is not the same $B_0$ as for the purely radial $p=2$ case, and in this combined case here the singularities combine at the origin. At that transition point the vortex at the origin has a topological charge of $2p+|l|=3$ since the singularities have merged. However, to add complication, away from $y=0$ the arrangement of the $2p+|l|$ singularities can change from being along $x$ to being separated along $t$, meaning that they trace more complicated $y$-oriented structure in space-time.

The complicated multidimensional character of the space-time singularity curves can be seen first in Figure~\ref{fig:5}c in the $y-t$ plane. With the smallest amount of spatial chirp in Fig.~\ref{fig:5}c.ii of $B=\sqrt{2}/4$, there is the linear curve corresponding to the central vortical singularity, and hyperbola-like curve corresponding to the two radial singularities. But clearly as the spatial chirp $B$ increases, this become more complicated and less intuitive, as seen in Fig.~\ref{fig:5}c.iii--iv. The vortical singularity curve going through the origin become nonlinear, and the hyperbola-like curves become displaced further from $t=0$. Although the plane cuts in Fig.~\ref{fig:5} show a significant amount of information, they still do not trace the singularities as they go off-plane in different dimensions.

\begin{figure}[htb]
\centering
\includegraphics[width=43mm]{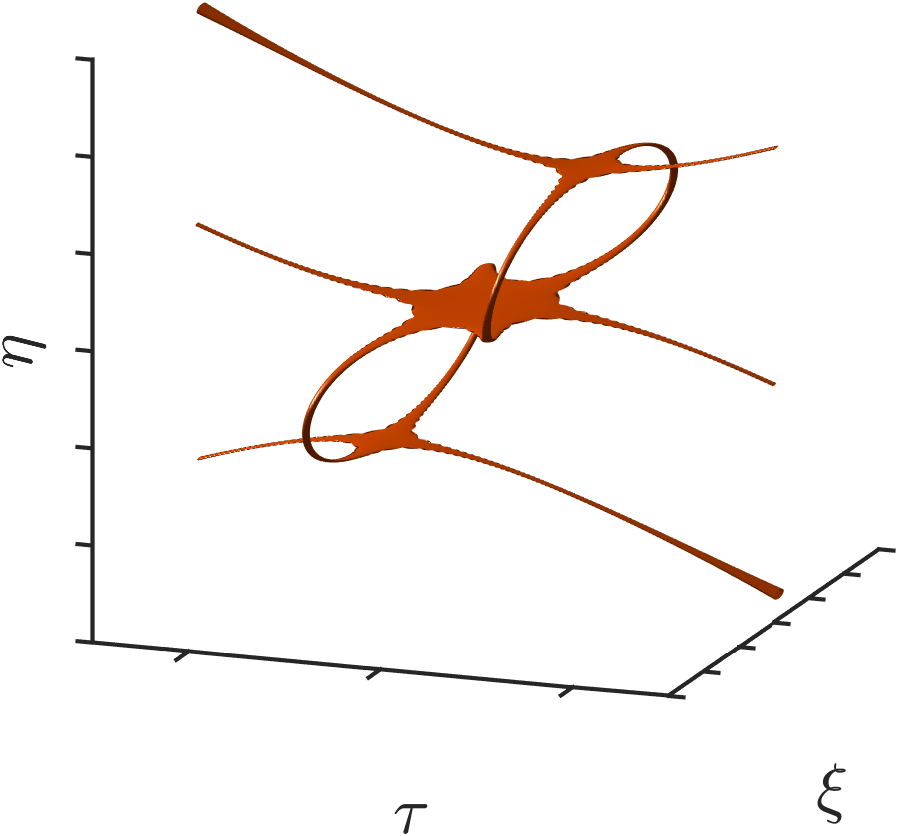}
\caption{Isosurface of the spatially-chirped LG beam with $l=1$ and $p=1$ at $z=0$ for a SC of $B=\sqrt{2}$. The range of the plot is $[-1.5,1.5]$ for all dimensions. The isosurface is at $1/5000$ of the peak intensity, such that the singularities are traced.}
\label{fig:6}
\end{figure}

To see this in a more complete manner, we can look at the singularity curves in 3D for one specific case in Figure~\ref{fig:6}. We show an isosurface at $1/5000$ of the peak intensity, which traces the areas surrounding the singularities in 3D space. We can see the triply charged singularity at the origin, and two doubly-charged singularities at off-plane positions at $\pm\xi$ and $\pm\tau$. All of those multi-charged point singularities, the two hyperbola-like curves, and the cubic-like curve are in the $\xi=0$ plane. But interestingly, between those three singular points, two ellipse-like curves in 3D are traced away from $\xi=0$. Unlike the $\{l,p\}=\{0,1\}$ case but similar to $\{l,p\}=\{0,2\}$, these curves are not in any plane, even a tilted plane, and truly trace a unique path.

Finally, to show one more example of a general LG beam with spatial chirp, we can manually find the solution to the integral when $l=\pm2$ and $p=1$ as

\begin{align}
\begin{split}
\frac{\mathcal{I}_{\pm2,1}}{\sqrt{\pi}}=&\pm c^2\left(c^2 - 3\right)+\frac{3\alpha^2}{2}-\frac{3\alpha^4}{4}\\
&\pm(2ic\alpha)\left(3-c^2-\frac{3\alpha^2}{2}\right)v\\
&+ 3(\alpha^2 - \alpha^4)v^2 \mp 2ic\alpha^3v^3-\alpha^4v^4.
\end{split}
\end{align}

\begin{figure}[htb]
\centering
\includegraphics[width=86mm]{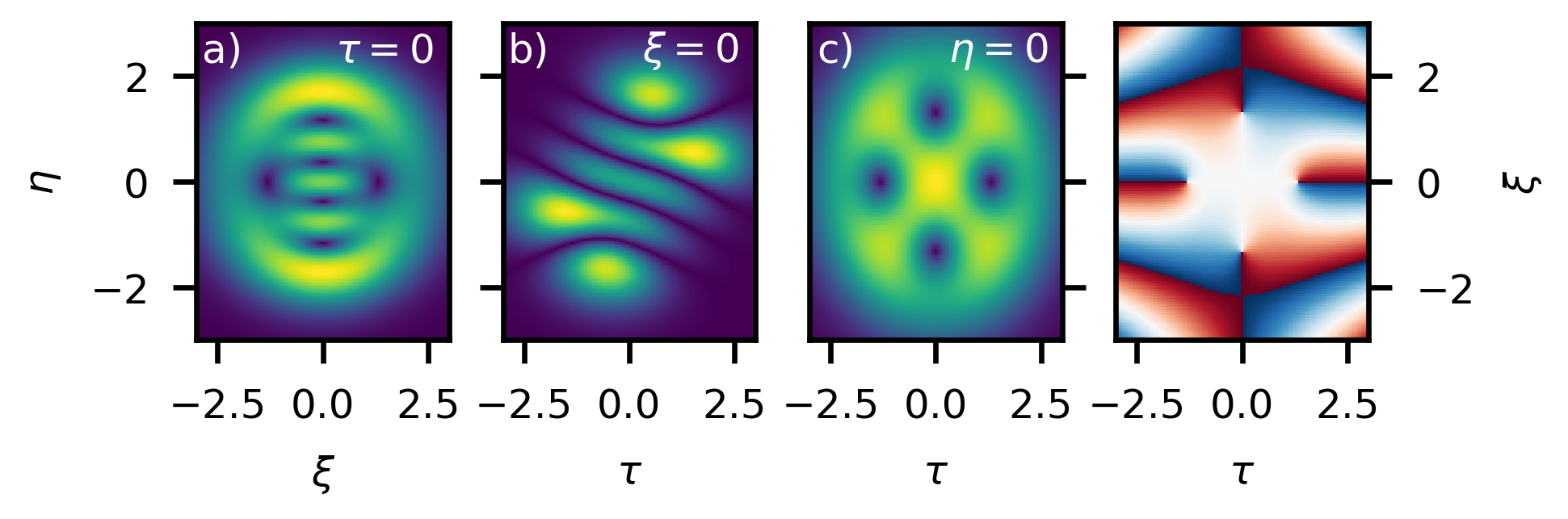}
\caption{The spatially-chirped LG beam with $l=2$ and $p=1$, at the focus ($z=0$) and with spatial chirp $B=1$. Amplitude profiles are shown for three plane cuts (a--c) with phase shown in the cut $y=0$ to the right of (c).} 
\label{fig:7}
\end{figure}

\noindent As expected, this polynomial solution for the integral is of order $2p+|l|=4$, which means that there are as many singularities with different paths in space-time. Without going into as much detail as for the previous cases, we can see in Figure~\ref{fig:7} one specific case. We first note that in the case of $|l|=2$, i.e. $|l|$ even, the singularity does not go through the origin (seen in Fig.~\ref{fig:7}a--b) as for the cases with $|l|$ odd. Looking at the integral solution, at $z=0$ ($f=1$) and with $B=1$ (and therefore $\alpha=1$) the singularities at $y=0$ are when $v^4=3/4$, which corresponds to the four equally-spaced points $\{\xi,\tau\}=\{0,\pm\sqrt[4]{3}\}$ and $\{\pm\sqrt[4]{3},0\}$ ($\sqrt[4]{3}\approx1.32$) seen in Fig.~\ref{fig:7}c. Beyond this, as for the previous cases, away from the plane cuts the singularities trace more complicated paths in space-time.

\subsection{Higher-order cases, propagation, and assumptions}

The previous sections used purely analytical tools to find the field profiles and singularity curves for different order LG beams with different amounts of spatial chirp. Essentially, although there is significant common behavior, all of the cases shown behaved differently due to the differing number, parity, and type of singularities present. To go beyond those examples, although it is tractable analytically, the order of the polynomials involved increases in both $v$ and $c$. And the experience of the low-order cases shows that most if not all of the terms in these mixed polynomials are non-zero. Even at $z=0$ ($f=1$) mixed-dimension polynomials of such high order no longer communicate intuitive behavior and regardless become cumbersome to write down and prone to needless errors. Therefore the practical way to calculate higher-order solutions than those presented in the previous sections is based on the numerical solution to the integral in Eq.~\ref{eq:LG_SC_simplified}.

Another increase in complexity that has been ignored until now is the behavior of LG beams with SC along propagation. At the focus ($z=0$ and $f=1$) it is significantly more simple mathematically and intuitively, since $f$ is purely real---this results in the singularities being in more well-defined positions even if that may vary based on the magnitude of $B$. In the far-field, i.e. far away from the focus, the situation is also relatively intuitive since $f$ becomes almost purely imaginary ($f\approx -i/\zeta$ when $\zeta\gg1$) and therefore the singularities are at very well-defined positions in $x$ and $t$ that are related to each other. This is a reflection of the fact that with significant propagation the SC will turn into PFT (just as focusing a pulse with PFT results in SC) and the singularities are just tilted versions of those of simple LG beams. Therefore the non-intuitive region, even for low-order beams, is in intermediate regions of propagation where $|\zeta|$ is comparable to unity.

\begin{figure}[htb]
\centering
\includegraphics[width=43mm]{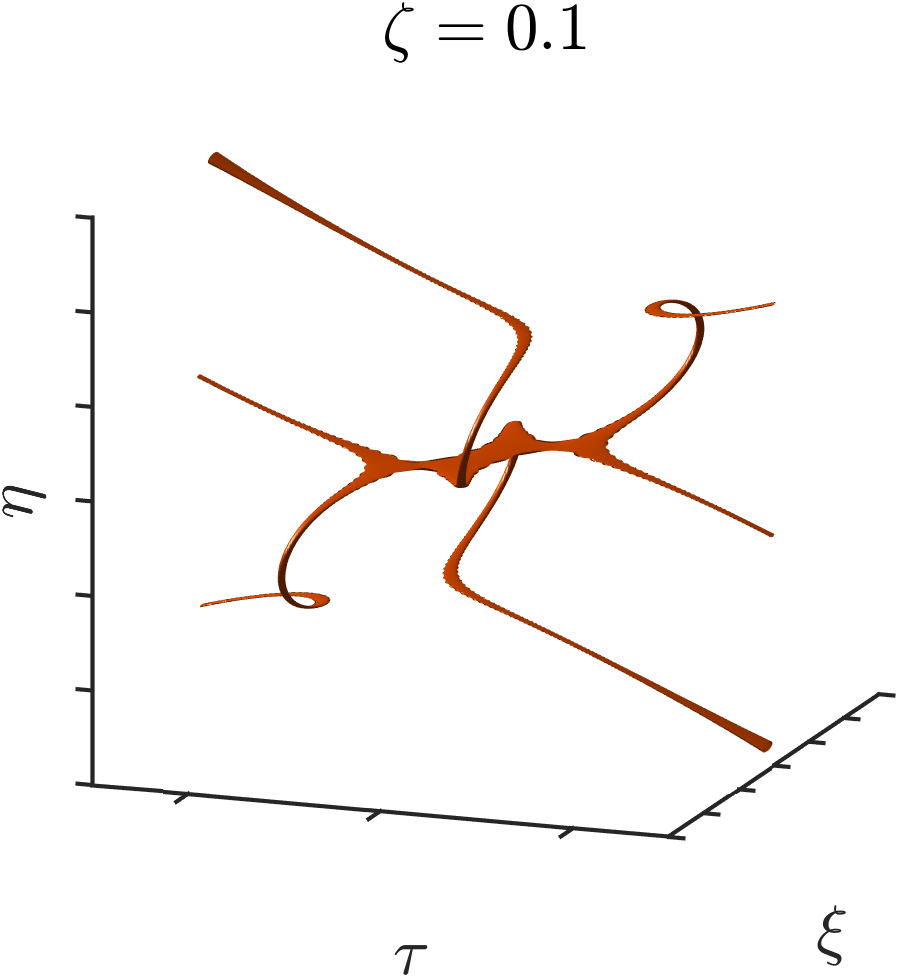}%
\includegraphics[width=43mm]{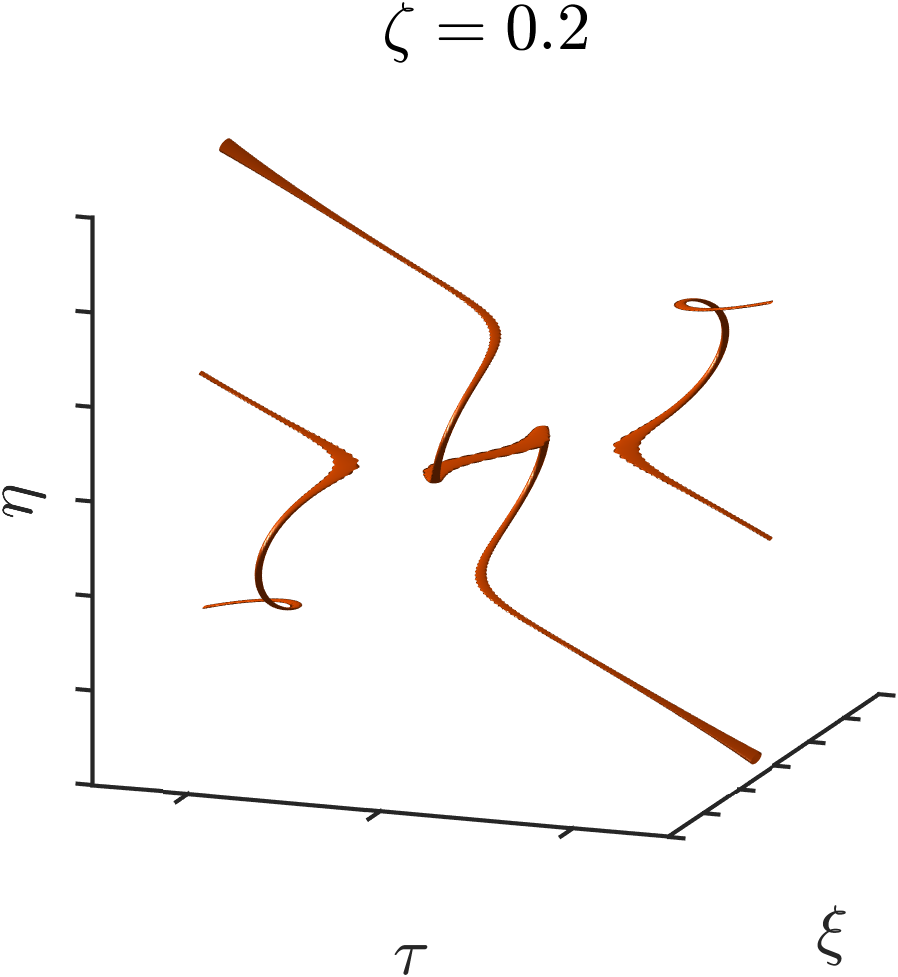}
\caption{Isosurface of the spatially-chirped LG beam with $l=1$ and $p=1$ at $\zeta=0.1$ (left) and 0.2 (right) for a SC of $B=\sqrt{2}$. The range of the plot is $[-1.5,1.5]$ for all dimensions. The isosurface is at $1/5000$ of the peak intensity, such that the singularities are traced.} 
\label{fig:8}
\end{figure}

In fact, the transition from SC to PFT is relatively straightforward, especially at $y=0$. What is much more complicated in this case is rather the position of the singularities in 3D space. It turns out, as seen in Fig.~\ref{fig:8} for the $\{l,p\}=\{1,1\}$ case that even at very small propagation distances away from the focus, $\zeta=0.1$ and 0.2, the singularity curves disconnect and warp away from each other in 3D space (compare to $z=0$ in Fig.~\ref{fig:7}). As $\zeta$ increases these singularities become more and more oriented along vertical lines and tilted in $x-t$ according to the PFT, and eventually also develop curvature due to the propagation and diffraction. If collected by a lens at a far-field position ($\zeta\gg1$), the line singularities become the circular plane singularities of a standard LG beam. Similar behavior occurs for other values of $l$ and $p$, i.e. the separation of merged singular points and warping towards PFT. The discontinuous behavior of the singularity curves and the merged singularities is therefore very highly localized around $z=0$.

Finally, while discussing complex aspects and nuances of this work, it is useful to revisit the assumptions that were made in deriving Eq.~\ref{eq:LG_SC_simplified}, which we used for all of the results that followed. The first assumption was that the Rayleigh range (and therefore $\zeta$ and $f$) does not depend on frequency. This indeed implies therefore that the beam size does depend on frequency, such that $w_0\propto1/\sqrt{\omega}$. This is the reason for the $\omega/\omega_0$ term in the exponent of Eq.~\ref{eq:Gauss_freq} that remains following that point. This assumption makes the mathematics that follows easier, since there is only one remaining extra $\omega$ term and it is in the exponential rather than in the denominator of a fraction. However, feasibly a laser source of a certain kind could have different properties, for example a fiber-based source where $w_0$ does not have any frequency dependence (and therefore $z_R$ does). Describing that case would require numerical integration as well. A secondary result of that remaining frequency term in Eq.~\ref{eq:Gauss_freq} is the term $\propto\rho^2$ in $t^{\prime\prime}$ (and $\tau$) that represents the time delay due to propagation and curvature. That aspect is a fully physical and important aspect of the propagation of focused pulsed light which was included in our analytical solutions.

A further assumption we make is that the pulse is not of few-cycle duration, i.e. $\Omega_0/\omega_0$ is small. Because of this we could ignore one term $\propto\xi\Omega^2$ and another term $\propto\Omega^3$ in Eq.~\ref{eq:LG_SC_freq} before eventually simplifying to Eqs.\ref{eq:LG_SC_integral}--\ref{eq:LG_SC_simplified}. Looking closer at the first of these terms, it can actually be followed through the analytic calculations. There is a correction in the quadratic part of the exponent of the form $\exp[(2fB\xi\Omega_0/\omega_0)(\Omega/\Omega_0)^2]$. This results in each term $1+fB^2$ being corrected to $1+fB^2-2fB\xi\Omega_0/\omega_0$, and $\xi$ being corrected to $\xi[1-2fB\xi\Omega_0/\omega_0]$ in $v$. The original argument was that in many-cycle pulses, $\omega_0\gg\Omega_0$, so these corrections can be ignored. Is this globally true, considering that the terms include both $\xi$ and $\zeta$ (via $f$)? Looking only at the magnitude of the correction term, since it becomes complex away from $z=0$, $|f|\xi$ is essentially $x/w(z)$ which means that the correction is only large at off-axis positions where there isn't significant intensity. The remaining part of the correction is $B$, meaning that the correction may also become significant if the spatial chirp $B$ is large, which we did not consider in this work.

Looking at the latter correction term, it takes the form $\exp[(B^2\Omega_0/\omega_0)(\Omega/\Omega_0)^3]$. Once again we used the justification that this is small when many cycle. However, we see that just like with the previous term, if the spatial chirp is large this is no longer necessarily a valid assumption. This term has no spatial dependence, meaning that it is a pure third-order dispersion and could be easily compensated for with standard optics. The previous term, however, is unavoidable in the case of large spatial chirp or large bandwidth (short duration). We suppose that the singularity trajectories could be modified, and that merged singularities may not perfectly merge. The assessment of these combined effects is not presented here and is outside the scope of this work, but can be done with the numerical solution of Eq.~\ref{eq:LG_SC_freq} without any simplifications.

\section{conclusion}

We have presented the analytical solutions for the electric field profile for Laguerre-Gaussian pulse beams with spatial chirp for a few different combinations of radial and vortical orders, and presented a recipe for numerically calculating for any order. We especially stress the position and curves of the singularities in 3D space-time, and discuss their differing form and behavior for the different cases shown, and how that changes depending on the magnitude of the spatial chirp. It is important to emphasize the difference in the work presented from space-time optical vortices with LG-like profiles in space-time~\cite{liuX24-2}---in this work the breaking of symmetry of the cylindrically-symmetric LG beam by the spatial chirp means that the resultant space-time profiles are not homogeneous or Gaussian-like in any dimension. The profiles and singularity curves that we presented, especially for the higher combined order $2p+|l|$, were truly complicated and unseparable in all three dimensions.

This work has built on previous works~\cite{porras23-3,jolly24-1,porras25-2} showing how the combination of simple paraxial propagation modes of free space and spatial chirp can result in complex and topological behavior of the field amplitude and the phase singularities. The case of Laguerre-Gaussian beams shown here is the most complex so far. This complexity provides a crucial context to more complicated scenarios that create more simple space-time behavior, expanding our knowledge of such behavior towards a goal of balancing fine control with simplicity (and fidelity and damage threshold). Applications for such fields remain relatively niche, such as attosecond optics, vacuum acceleration or manipulation of electrons, or ultrafast pulsed versions of information transfer, optical tweezers, or even remote sensing. It is also promising to transfer such concepts outside of the optical frequency range where more applications may be available.

\section*{Funding}
Fonds De La Recherche Scientifique - FNRS.

\section*{Disclosures}
The authors declare no conflicts of interest.

\section*{Data availability}
Data underlying the results presented in this paper are not publicly available at this time but may be obtained from the authors upon reasonable request.


%


\begin{thebibliography}{31}%
	\makeatletter
	\providecommand \@ifxundefined [1]{%
		\@ifx{#1\undefined}
	}%
	\providecommand \@ifnum [1]{%
		\ifnum #1\expandafter \@firstoftwo
		\else \expandafter \@secondoftwo
		\fi
	}%
	\providecommand \@ifx [1]{%
		\ifx #1\expandafter \@firstoftwo
		\else \expandafter \@secondoftwo
		\fi
	}%
	\providecommand \natexlab [1]{#1}%
	\providecommand \enquote  [1]{``#1''}%
	\providecommand \bibnamefont  [1]{#1}%
	\providecommand \bibfnamefont [1]{#1}%
	\providecommand \citenamefont [1]{#1}%
	\providecommand \href@noop [0]{\@secondoftwo}%
	\providecommand \href [0]{\begingroup \@sanitize@url \@href}%
	\providecommand \@href[1]{\@@startlink{#1}\@@href}%
	\providecommand \@@href[1]{\endgroup#1\@@endlink}%
	\providecommand \@sanitize@url [0]{\catcode `\\12\catcode `\$12\catcode
		`\&12\catcode `\#12\catcode `\^12\catcode `\_12\catcode `\%12\relax}%
	\providecommand \@@startlink[1]{}%
	\providecommand \@@endlink[0]{}%
	\providecommand \url  [0]{\begingroup\@sanitize@url \@url }%
	\providecommand \@url [1]{\endgroup\@href {#1}{\urlprefix }}%
	\providecommand \urlprefix  [0]{URL }%
	\providecommand \Eprint [0]{\href }%
	\providecommand \doibase [0]{https://doi.org/}%
	\providecommand \selectlanguage [0]{\@gobble}%
	\providecommand \bibinfo  [0]{\@secondoftwo}%
	\providecommand \bibfield  [0]{\@secondoftwo}%
	\providecommand \translation [1]{[#1]}%
	\providecommand \BibitemOpen [0]{}%
	\providecommand \bibitemStop [0]{}%
	\providecommand \bibitemNoStop [0]{.\EOS\space}%
	\providecommand \EOS [0]{\spacefactor3000\relax}%
	\providecommand \BibitemShut  [1]{\csname bibitem#1\endcsname}%
	\let\auto@bib@innerbib\@empty
	\bibitem [{\citenamefont {Akturk}\ \emph {et~al.}(2005)\citenamefont {Akturk},
		\citenamefont {Gu}, \citenamefont {Gabolde},\ and\ \citenamefont
		{Trebino}}]{akturk05}%
	\BibitemOpen
	\bibfield  {author} {\bibinfo {author} {\bibfnamefont {S.}~\bibnamefont
			{Akturk}}, \bibinfo {author} {\bibfnamefont {X.}~\bibnamefont {Gu}}, \bibinfo
		{author} {\bibfnamefont {P.}~\bibnamefont {Gabolde}},\ and\ \bibinfo {author}
		{\bibfnamefont {R.}~\bibnamefont {Trebino}},\ }\href@noop {} {\bibfield
		{journal} {\bibinfo  {journal} {Optics Express}\ }\textbf {\bibinfo {volume}
			{13}},\ \bibinfo {pages} {8642} (\bibinfo {year} {2005})}\BibitemShut
	{NoStop}%
	\bibitem [{\citenamefont {Akturk}\ \emph {et~al.}(2010)\citenamefont {Akturk},
		\citenamefont {Gu}, \citenamefont {Bowlan},\ and\ \citenamefont
		{Trebino}}]{akturk10}%
	\BibitemOpen
	\bibfield  {author} {\bibinfo {author} {\bibfnamefont {S.}~\bibnamefont
			{Akturk}}, \bibinfo {author} {\bibfnamefont {X.}~\bibnamefont {Gu}}, \bibinfo
		{author} {\bibfnamefont {P.}~\bibnamefont {Bowlan}},\ and\ \bibinfo {author}
		{\bibfnamefont {R.}~\bibnamefont {Trebino}},\ }\href@noop {} {\bibfield
		{journal} {\bibinfo  {journal} {Journal of Optics}\ }\textbf {\bibinfo
			{volume} {12}},\ \bibinfo {pages} {093001} (\bibinfo {year}
		{2010})}\BibitemShut {NoStop}%
	\bibitem [{\citenamefont {Shen}\ \emph {et~al.}(2023)\citenamefont {Shen},
		\citenamefont {Zhan}, \citenamefont {Wright}, \citenamefont
		{Christodoulides}, \citenamefont {Wise}, \citenamefont {Willner},
		\citenamefont {heng Zou}, \citenamefont {Zhao}, \citenamefont {Porras},
		\citenamefont {Chong}, \citenamefont {Wan}, \citenamefont {Bliokh},
		\citenamefont {Liao}, \citenamefont {Hernández-García}, \citenamefont
		{Murnane}, \citenamefont {Yessenov}, \citenamefont {Abouraddy}, \citenamefont
		{Wong}, \citenamefont {Go}, \citenamefont {Kumar}, \citenamefont {Guo},
		\citenamefont {Fan}, \citenamefont {Papasimakis}, \citenamefont {Zheludev},
		\citenamefont {Chen}, \citenamefont {Zhu}, \citenamefont {Agrawal},
		\citenamefont {Mounaix}, \citenamefont {Fontaine}, \citenamefont {Carpenter},
		\citenamefont {Jolly}, \citenamefont {Dorrer}, \citenamefont {Alonso},
		\citenamefont {Lopez-Quintas}, \citenamefont {López-Ripa}, \citenamefont
		{Íñigo J.~Sola}, \citenamefont {Huang}, \citenamefont {Zhang},
		\citenamefont {Ruan}, \citenamefont {Dorrah}, \citenamefont {Capasso},\ and\
		\citenamefont {Forbes}}]{shenY23}%
	\BibitemOpen
	\bibfield  {author} {\bibinfo {author} {\bibfnamefont {Y.}~\bibnamefont
			{Shen}}, \bibinfo {author} {\bibfnamefont {Q.}~\bibnamefont {Zhan}}, \bibinfo
		{author} {\bibfnamefont {L.~G.}\ \bibnamefont {Wright}}, \bibinfo {author}
		{\bibfnamefont {D.~N.}\ \bibnamefont {Christodoulides}}, \bibinfo {author}
		{\bibfnamefont {F.~W.}\ \bibnamefont {Wise}}, \bibinfo {author}
		{\bibfnamefont {A.~E.}\ \bibnamefont {Willner}}, \bibinfo {author}
		{\bibfnamefont {K.}~\bibnamefont {heng Zou}}, \bibinfo {author}
		{\bibfnamefont {Z.}~\bibnamefont {Zhao}}, \bibinfo {author} {\bibfnamefont
			{M.~A.}\ \bibnamefont {Porras}}, \bibinfo {author} {\bibfnamefont
			{A.}~\bibnamefont {Chong}}, \bibinfo {author} {\bibfnamefont
			{C.}~\bibnamefont {Wan}}, \bibinfo {author} {\bibfnamefont {K.~Y.}\
			\bibnamefont {Bliokh}}, \bibinfo {author} {\bibfnamefont {C.-T.}\
			\bibnamefont {Liao}}, \bibinfo {author} {\bibfnamefont {C.}~\bibnamefont
			{Hernández-García}}, \bibinfo {author} {\bibfnamefont {M.~M.}\ \bibnamefont
			{Murnane}}, \bibinfo {author} {\bibfnamefont {M.}~\bibnamefont {Yessenov}},
		\bibinfo {author} {\bibfnamefont {A.~F.}\ \bibnamefont {Abouraddy}}, \bibinfo
		{author} {\bibfnamefont {L.~J.}\ \bibnamefont {Wong}}, \bibinfo {author}
		{\bibfnamefont {M.}~\bibnamefont {Go}}, \bibinfo {author} {\bibfnamefont
			{S.}~\bibnamefont {Kumar}}, \bibinfo {author} {\bibfnamefont
			{C.}~\bibnamefont {Guo}}, \bibinfo {author} {\bibfnamefont {S.}~\bibnamefont
			{Fan}}, \bibinfo {author} {\bibfnamefont {N.}~\bibnamefont {Papasimakis}},
		\bibinfo {author} {\bibfnamefont {N.~I.}\ \bibnamefont {Zheludev}}, \bibinfo
		{author} {\bibfnamefont {L.}~\bibnamefont {Chen}}, \bibinfo {author}
		{\bibfnamefont {W.}~\bibnamefont {Zhu}}, \bibinfo {author} {\bibfnamefont
			{A.}~\bibnamefont {Agrawal}}, \bibinfo {author} {\bibfnamefont
			{M.}~\bibnamefont {Mounaix}}, \bibinfo {author} {\bibfnamefont {N.~K.}\
			\bibnamefont {Fontaine}}, \bibinfo {author} {\bibfnamefont {J.}~\bibnamefont
			{Carpenter}}, \bibinfo {author} {\bibfnamefont {S.~W.}\ \bibnamefont
			{Jolly}}, \bibinfo {author} {\bibfnamefont {C.}~\bibnamefont {Dorrer}},
		\bibinfo {author} {\bibfnamefont {B.}~\bibnamefont {Alonso}}, \bibinfo
		{author} {\bibfnamefont {I.}~\bibnamefont {Lopez-Quintas}}, \bibinfo {author}
		{\bibfnamefont {M.}~\bibnamefont {López-Ripa}}, \bibinfo {author}
		{\bibnamefont {Íñigo J.~Sola}}, \bibinfo {author} {\bibfnamefont
			{J.}~\bibnamefont {Huang}}, \bibinfo {author} {\bibfnamefont
			{H.}~\bibnamefont {Zhang}}, \bibinfo {author} {\bibfnamefont
			{Z.}~\bibnamefont {Ruan}}, \bibinfo {author} {\bibfnamefont {A.~H.}\
			\bibnamefont {Dorrah}}, \bibinfo {author} {\bibfnamefont {F.}~\bibnamefont
			{Capasso}},\ and\ \bibinfo {author} {\bibfnamefont {A.}~\bibnamefont
			{Forbes}},\ }\href@noop {} {\bibfield  {journal} {\bibinfo  {journal}
			{Journal of optics}\ }\textbf {\bibinfo {volume} {25}},\ \bibinfo {pages}
		{093001} (\bibinfo {year} {2023})}\BibitemShut {NoStop}%
	\bibitem [{\citenamefont {Liu}\ \emph {et~al.}(2024{\natexlab{a}})\citenamefont
		{Liu}, \citenamefont {Cao},\ and\ \citenamefont {Zhan}}]{liuX24}%
	\BibitemOpen
	\bibfield  {author} {\bibinfo {author} {\bibfnamefont {X.}~\bibnamefont
			{Liu}}, \bibinfo {author} {\bibfnamefont {Q.}~\bibnamefont {Cao}},\ and\
		\bibinfo {author} {\bibfnamefont {Q.}~\bibnamefont {Zhan}},\ }\href@noop {}
	{\bibfield  {journal} {\bibinfo  {journal} {Photonics Insights}\ }\textbf
		{\bibinfo {volume} {3}},\ \bibinfo {pages} {R08} (\bibinfo {year}
		{2024}{\natexlab{a}})}\BibitemShut {NoStop}%
	\bibitem [{\citenamefont {Mounaix}\ \emph {et~al.}(2020)\citenamefont
		{Mounaix}, \citenamefont {Fontaine}, \citenamefont {Neilson}, \citenamefont
		{Ryf}, \citenamefont {Chen}, \citenamefont {Alvarado-Zacarias},\ and\
		\citenamefont {Carpenter}}]{mounaix20}%
	\BibitemOpen
	\bibfield  {author} {\bibinfo {author} {\bibfnamefont {M.}~\bibnamefont
			{Mounaix}}, \bibinfo {author} {\bibfnamefont {N.~K.}\ \bibnamefont
			{Fontaine}}, \bibinfo {author} {\bibfnamefont {D.~T.}\ \bibnamefont
			{Neilson}}, \bibinfo {author} {\bibfnamefont {R.}~\bibnamefont {Ryf}},
		\bibinfo {author} {\bibfnamefont {H.}~\bibnamefont {Chen}}, \bibinfo {author}
		{\bibfnamefont {J.~C.}\ \bibnamefont {Alvarado-Zacarias}},\ and\ \bibinfo
		{author} {\bibfnamefont {J.}~\bibnamefont {Carpenter}},\ }\href@noop {}
	{\bibfield  {journal} {\bibinfo  {journal} {Nature Communications}\ }\textbf
		{\bibinfo {volume} {11}},\ \bibinfo {pages} {5813} (\bibinfo {year}
		{2020})}\BibitemShut {NoStop}%
	\bibitem [{\citenamefont {Sukhorukov}\ and\ \citenamefont
		{Yangirova}(2005)}]{sukhorukov05}%
	\BibitemOpen
	\bibfield  {author} {\bibinfo {author} {\bibfnamefont {A.~P.}\ \bibnamefont
			{Sukhorukov}}\ and\ \bibinfo {author} {\bibfnamefont {V.~V.}\ \bibnamefont
			{Yangirova}},\ }in\ \href@noop {} {\emph {\bibinfo {booktitle} {Nonlinear
				Optics Applications}}},\ Vol.\ \bibinfo {volume} {5949},\ \bibinfo
	{organization} {International Society for Optics and Photonics}\ (\bibinfo
	{publisher} {SPIE},\ \bibinfo {year} {2005})\ p.\ \bibinfo {pages}
	{594906}\BibitemShut {NoStop}%
	\bibitem [{\citenamefont {Jhajj}\ \emph {et~al.}(2016)\citenamefont {Jhajj},
		\citenamefont {Larkin}, \citenamefont {Rosenthal}, \citenamefont {Zahedpour},
		\citenamefont {Wahlstrand},\ and\ \citenamefont {Milchberg}}]{jhajj16}%
	\BibitemOpen
	\bibfield  {author} {\bibinfo {author} {\bibfnamefont {N.}~\bibnamefont
			{Jhajj}}, \bibinfo {author} {\bibfnamefont {I.}~\bibnamefont {Larkin}},
		\bibinfo {author} {\bibfnamefont {E.~W.}\ \bibnamefont {Rosenthal}}, \bibinfo
		{author} {\bibfnamefont {S.}~\bibnamefont {Zahedpour}}, \bibinfo {author}
		{\bibfnamefont {J.~K.}\ \bibnamefont {Wahlstrand}},\ and\ \bibinfo {author}
		{\bibfnamefont {H.~M.}\ \bibnamefont {Milchberg}},\ }\href@noop {} {\bibfield
		{journal} {\bibinfo  {journal} {Physical Review X}\ }\textbf {\bibinfo
			{volume} {6}},\ \bibinfo {pages} {031037} (\bibinfo {year}
		{2016})}\BibitemShut {NoStop}%
	\bibitem [{\citenamefont {Wan}\ \emph {et~al.}(2023)\citenamefont {Wan},
		\citenamefont {Chong},\ and\ \citenamefont {Zhan}}]{wanC23}%
	\BibitemOpen
	\bibfield  {author} {\bibinfo {author} {\bibfnamefont {C.}~\bibnamefont
			{Wan}}, \bibinfo {author} {\bibfnamefont {A.}~\bibnamefont {Chong}},\ and\
		\bibinfo {author} {\bibfnamefont {Q.}~\bibnamefont {Zhan}},\ }\href@noop {}
	{\bibfield  {journal} {\bibinfo  {journal} {eLight}\ }\textbf {\bibinfo
			{volume} {3}} (\bibinfo {year} {2023})}\BibitemShut {NoStop}%
	\bibitem [{\citenamefont {Porras}(2023)}]{porras23-1}%
	\BibitemOpen
	\bibfield  {author} {\bibinfo {author} {\bibfnamefont {M.~A.}\ \bibnamefont
			{Porras}},\ }\href@noop {} {\bibfield  {journal} {\bibinfo  {journal} {Optics
				Letters}\ }\textbf {\bibinfo {volume} {48}},\ \bibinfo {pages} {367}
		(\bibinfo {year} {2023})}\BibitemShut {NoStop}%
	\bibitem [{\citenamefont {Bekshaev}(2024)}]{bekshaev2024spatiotemporal}%
	\BibitemOpen
	\bibfield  {author} {\bibinfo {author} {\bibfnamefont {A.}~\bibnamefont
			{Bekshaev}},\ }\href@noop {} {\bibfield  {journal} {\bibinfo  {journal} {APL
				Photonics}\ }\textbf {\bibinfo {volume} {9}} (\bibinfo {year}
		{2024})}\BibitemShut {NoStop}%
	\bibitem [{\citenamefont {Chong}\ \emph {et~al.}(2020)\citenamefont {Chong},
		\citenamefont {Wan}, \citenamefont {Chen},\ and\ \citenamefont
		{Zhan}}]{chong20}%
	\BibitemOpen
	\bibfield  {author} {\bibinfo {author} {\bibfnamefont {A.}~\bibnamefont
			{Chong}}, \bibinfo {author} {\bibfnamefont {C.}~\bibnamefont {Wan}}, \bibinfo
		{author} {\bibfnamefont {J.}~\bibnamefont {Chen}},\ and\ \bibinfo {author}
		{\bibfnamefont {Q.}~\bibnamefont {Zhan}},\ }\href@noop {} {\bibfield
		{journal} {\bibinfo  {journal} {Nature Photonics}\ }\textbf {\bibinfo
			{volume} {14}},\ \bibinfo {pages} {350} (\bibinfo {year} {2020})}\BibitemShut
	{NoStop}%
	\bibitem [{\citenamefont {Wang}\ \emph {et~al.}(2021)\citenamefont {Wang},
		\citenamefont {Guo}, \citenamefont {Jin}, \citenamefont {Song},\ and\
		\citenamefont {Fan}}]{wangH21}%
	\BibitemOpen
	\bibfield  {author} {\bibinfo {author} {\bibfnamefont {H.}~\bibnamefont
			{Wang}}, \bibinfo {author} {\bibfnamefont {C.}~\bibnamefont {Guo}}, \bibinfo
		{author} {\bibfnamefont {W.}~\bibnamefont {Jin}}, \bibinfo {author}
		{\bibfnamefont {A.~Y.}\ \bibnamefont {Song}},\ and\ \bibinfo {author}
		{\bibfnamefont {S.}~\bibnamefont {Fan}},\ }\href@noop {} {\bibfield
		{journal} {\bibinfo  {journal} {Optica}\ }\textbf {\bibinfo {volume} {8}},\
		\bibinfo {pages} {966} (\bibinfo {year} {2021})}\BibitemShut {NoStop}%
	\bibitem [{\citenamefont {Zang}\ \emph {et~al.}(2022)\citenamefont {Zang},
		\citenamefont {Mirando},\ and\ \citenamefont {Chong}}]{zangY22}%
	\BibitemOpen
	\bibfield  {author} {\bibinfo {author} {\bibfnamefont {Y.}~\bibnamefont
			{Zang}}, \bibinfo {author} {\bibfnamefont {A.}~\bibnamefont {Mirando}},\ and\
		\bibinfo {author} {\bibfnamefont {A.}~\bibnamefont {Chong}},\ }\href@noop {}
	{\bibfield  {journal} {\bibinfo  {journal} {Nanophotonics}\ }\textbf
		{\bibinfo {volume} {11}},\ \bibinfo {pages} {745} (\bibinfo {year}
		{2022})}\BibitemShut {NoStop}%
	\bibitem [{\citenamefont {Liu}\ \emph {et~al.}(2024{\natexlab{b}})\citenamefont
		{Liu}, \citenamefont {Cao}, \citenamefont {Zhang}, \citenamefont {Chong},
		\citenamefont {Cai},\ and\ \citenamefont {Zhan}}]{liuX24-2}%
	\BibitemOpen
	\bibfield  {author} {\bibinfo {author} {\bibfnamefont {X.}~\bibnamefont
			{Liu}}, \bibinfo {author} {\bibfnamefont {Q.}~\bibnamefont {Cao}}, \bibinfo
		{author} {\bibfnamefont {N.}~\bibnamefont {Zhang}}, \bibinfo {author}
		{\bibfnamefont {A.}~\bibnamefont {Chong}}, \bibinfo {author} {\bibfnamefont
			{Y.}~\bibnamefont {Cai}},\ and\ \bibinfo {author} {\bibfnamefont
			{Q.}~\bibnamefont {Zhan}},\ }\href@noop {} {\bibfield  {journal} {\bibinfo
			{journal} {Nature Communications}\ }\textbf {\bibinfo {volume} {15}},\
		\bibinfo {pages} {5435} (\bibinfo {year} {2024}{\natexlab{b}})}\BibitemShut
	{NoStop}%
	\bibitem [{\citenamefont {Chen}\ \emph {et~al.}(2022)\citenamefont {Chen},
		\citenamefont {Zhang}, \citenamefont {Liu}, \citenamefont {Meng},
		\citenamefont {Dudley},\ and\ \citenamefont {Lu}}]{chenW22}%
	\BibitemOpen
	\bibfield  {author} {\bibinfo {author} {\bibfnamefont {W.}~\bibnamefont
			{Chen}}, \bibinfo {author} {\bibfnamefont {W.}~\bibnamefont {Zhang}},
		\bibinfo {author} {\bibfnamefont {Y.}~\bibnamefont {Liu}}, \bibinfo {author}
		{\bibfnamefont {F.-C.}\ \bibnamefont {Meng}}, \bibinfo {author}
		{\bibfnamefont {J.~M.}\ \bibnamefont {Dudley}},\ and\ \bibinfo {author}
		{\bibfnamefont {Y.-Q.}\ \bibnamefont {Lu}},\ }\href@noop {} {\bibfield
		{journal} {\bibinfo  {journal} {Nature Communications}\ }\textbf {\bibinfo
			{volume} {13}},\ \bibinfo {pages} {4021} (\bibinfo {year}
		{2022})}\BibitemShut {NoStop}%
	\bibitem [{\citenamefont {Cao}\ \emph {et~al.}(2024)\citenamefont {Cao},
		\citenamefont {Zhang}, \citenamefont {Chong},\ and\ \citenamefont
		{Zhan}}]{caoQ24}%
	\BibitemOpen
	\bibfield  {author} {\bibinfo {author} {\bibfnamefont {Q.}~\bibnamefont
			{Cao}}, \bibinfo {author} {\bibfnamefont {N.}~\bibnamefont {Zhang}}, \bibinfo
		{author} {\bibfnamefont {A.}~\bibnamefont {Chong}},\ and\ \bibinfo {author}
		{\bibfnamefont {Q.}~\bibnamefont {Zhan}},\ }\href@noop {} {\bibfield
		{journal} {\bibinfo  {journal} {Nature Communications}\ }\textbf {\bibinfo
			{volume} {15}},\ \bibinfo {pages} {7821} (\bibinfo {year}
		{2024})}\BibitemShut {NoStop}%
	\bibitem [{\citenamefont {Gu}\ \emph {et~al.}(2004)\citenamefont {Gu},
		\citenamefont {Akturk},\ and\ \citenamefont {Trebino}}]{gu04}%
	\BibitemOpen
	\bibfield  {author} {\bibinfo {author} {\bibfnamefont {X.}~\bibnamefont
			{Gu}}, \bibinfo {author} {\bibfnamefont {S.}~\bibnamefont {Akturk}},\ and\
		\bibinfo {author} {\bibfnamefont {R.}~\bibnamefont {Trebino}},\ }\href@noop
	{} {\bibfield  {journal} {\bibinfo  {journal} {Optics Communications}\
		}\textbf {\bibinfo {volume} {242}},\ \bibinfo {pages} {599} (\bibinfo {year}
		{2004})}\BibitemShut {NoStop}%
	\bibitem [{\citenamefont {Vincenti}\ and\ \citenamefont
		{Qu{\'e}r{\'e}}(2012)}]{vincenti12}%
	\BibitemOpen
	\bibfield  {author} {\bibinfo {author} {\bibfnamefont {H.}~\bibnamefont
			{Vincenti}}\ and\ \bibinfo {author} {\bibfnamefont {F.}~\bibnamefont
			{Qu{\'e}r{\'e}}},\ }\href@noop {} {\bibfield  {journal} {\bibinfo  {journal}
			{Physical Review Letters}\ }\textbf {\bibinfo {volume} {108}},\ \bibinfo
		{pages} {113904} (\bibinfo {year} {2012})}\BibitemShut {NoStop}%
	\bibitem [{\citenamefont {Wheeler}\ \emph {et~al.}(2012)\citenamefont
		{Wheeler}, \citenamefont {Borot}, \citenamefont {Monchoce}, \citenamefont
		{Vincenti}, \citenamefont {Ricci}, \citenamefont {Malvache}, \citenamefont
		{Lopez-Martens},\ and\ \citenamefont {Qu{\'e}r{\'e}}}]{wheeler12}%
	\BibitemOpen
	\bibfield  {author} {\bibinfo {author} {\bibfnamefont {J.~A.}\ \bibnamefont
			{Wheeler}}, \bibinfo {author} {\bibfnamefont {A.}~\bibnamefont {Borot}},
		\bibinfo {author} {\bibfnamefont {S.}~\bibnamefont {Monchoce}}, \bibinfo
		{author} {\bibfnamefont {H.}~\bibnamefont {Vincenti}}, \bibinfo {author}
		{\bibfnamefont {A.}~\bibnamefont {Ricci}}, \bibinfo {author} {\bibfnamefont
			{A.}~\bibnamefont {Malvache}}, \bibinfo {author} {\bibfnamefont
			{R.}~\bibnamefont {Lopez-Martens}},\ and\ \bibinfo {author} {\bibfnamefont
			{F.}~\bibnamefont {Qu{\'e}r{\'e}}},\ }\href
	{https://doi.org/10.1038/NPHOTON.2012.284} {\bibfield  {journal} {\bibinfo
			{journal} {Nature Photonics}\ }\textbf {\bibinfo {volume} {6}},\ \bibinfo
		{pages} {829} (\bibinfo {year} {2012})}\BibitemShut {NoStop}%
	\bibitem [{\citenamefont {Kim}\ \emph {et~al.}(2013)\citenamefont {Kim},
		\citenamefont {Zhang}, \citenamefont {Ruchon}, \citenamefont {Hergott},
		\citenamefont {Auguste}, \citenamefont {Villeneuve}, \citenamefont {Corkum},\
		and\ \citenamefont {Qu{\'e}r{\'e}}}]{kim13-2}%
	\BibitemOpen
	\bibfield  {author} {\bibinfo {author} {\bibfnamefont {K.~T.}\ \bibnamefont
			{Kim}}, \bibinfo {author} {\bibfnamefont {C.}~\bibnamefont {Zhang}}, \bibinfo
		{author} {\bibfnamefont {T.}~\bibnamefont {Ruchon}}, \bibinfo {author}
		{\bibfnamefont {J.-F.}\ \bibnamefont {Hergott}}, \bibinfo {author}
		{\bibfnamefont {T.}~\bibnamefont {Auguste}}, \bibinfo {author} {\bibfnamefont
			{D.~M.}\ \bibnamefont {Villeneuve}}, \bibinfo {author} {\bibfnamefont
			{P.~B.}\ \bibnamefont {Corkum}},\ and\ \bibinfo {author} {\bibfnamefont
			{F.}~\bibnamefont {Qu{\'e}r{\'e}}},\ }\href
	{https://doi.org/10.1038/NPHOTON.2013.170} {\bibfield  {journal} {\bibinfo
			{journal} {Nature Photonics}\ }\textbf {\bibinfo {volume} {7}},\ \bibinfo
		{pages} {651} (\bibinfo {year} {2013})}\BibitemShut {NoStop}%
	\bibitem [{\citenamefont {Qu{\'e}r{\'e}}\ \emph {et~al.}(2014)\citenamefont
		{Qu{\'e}r{\'e}}, \citenamefont {Vincenti}, \citenamefont {Borot},
		\citenamefont {Monchoc{\'e}}, \citenamefont {Hammond}, \citenamefont {Kim},
		\citenamefont {Wheeler}, \citenamefont {Zhang}, \citenamefont {Ruchon},
		\citenamefont {Auguste}, \citenamefont {Hergott}, \citenamefont {Villeneuve},
		\citenamefont {Corkum},\ and\ \citenamefont {Lopez-Martens}}]{quere14}%
	\BibitemOpen
	\bibfield  {author} {\bibinfo {author} {\bibfnamefont {F.}~\bibnamefont
			{Qu{\'e}r{\'e}}}, \bibinfo {author} {\bibfnamefont {H.}~\bibnamefont
			{Vincenti}}, \bibinfo {author} {\bibfnamefont {A.}~\bibnamefont {Borot}},
		\bibinfo {author} {\bibfnamefont {S.}~\bibnamefont {Monchoc{\'e}}}, \bibinfo
		{author} {\bibfnamefont {T.~J.}\ \bibnamefont {Hammond}}, \bibinfo {author}
		{\bibfnamefont {K.~T.}\ \bibnamefont {Kim}}, \bibinfo {author} {\bibfnamefont
			{J.~A.}\ \bibnamefont {Wheeler}}, \bibinfo {author} {\bibfnamefont
			{C.}~\bibnamefont {Zhang}}, \bibinfo {author} {\bibfnamefont
			{T.}~\bibnamefont {Ruchon}}, \bibinfo {author} {\bibfnamefont
			{T.}~\bibnamefont {Auguste}}, \bibinfo {author} {\bibfnamefont {J.~F.}\
			\bibnamefont {Hergott}}, \bibinfo {author} {\bibfnamefont {D.~M.}\
			\bibnamefont {Villeneuve}}, \bibinfo {author} {\bibfnamefont {P.~B.}\
			\bibnamefont {Corkum}},\ and\ \bibinfo {author} {\bibfnamefont
			{R.}~\bibnamefont {Lopez-Martens}},\ }\href@noop {} {\bibfield  {journal}
		{\bibinfo  {journal} {Journal of Physics B: Atomic, Molecular and Optical
				Physics}\ }\textbf {\bibinfo {volume} {47}},\ \bibinfo {pages} {124004}
		(\bibinfo {year} {2014})}\BibitemShut {NoStop}%
	\bibitem [{\citenamefont {Auguste}\ \emph {et~al.}(2016)\citenamefont
		{Auguste}, \citenamefont {Gobert}, \citenamefont {Ruchon},\ and\
		\citenamefont {Qu{\'e}r{\'e}}}]{auguste16}%
	\BibitemOpen
	\bibfield  {author} {\bibinfo {author} {\bibfnamefont {T.}~\bibnamefont
			{Auguste}}, \bibinfo {author} {\bibfnamefont {O.}~\bibnamefont {Gobert}},
		\bibinfo {author} {\bibfnamefont {T.}~\bibnamefont {Ruchon}},\ and\ \bibinfo
		{author} {\bibfnamefont {F.}~\bibnamefont {Qu{\'e}r{\'e}}},\ }\href@noop {}
	{\bibfield  {journal} {\bibinfo  {journal} {Physical Review A}\ }\textbf
		{\bibinfo {volume} {93}},\ \bibinfo {pages} {033825} (\bibinfo {year}
		{2016})}\BibitemShut {NoStop}%
	\bibitem [{\citenamefont {Porras}\ and\ \citenamefont
		{Jolly}(2023)}]{porras23-3}%
	\BibitemOpen
	\bibfield  {author} {\bibinfo {author} {\bibfnamefont {M.~A.}\ \bibnamefont
			{Porras}}\ and\ \bibinfo {author} {\bibfnamefont {S.~W.}\ \bibnamefont
			{Jolly}},\ }\href@noop {} {\bibfield  {journal} {\bibinfo  {journal} {Optics
				Letters}\ }\textbf {\bibinfo {volume} {48}},\ \bibinfo {pages} {6448}
		(\bibinfo {year} {2023})}\BibitemShut {NoStop}%
	\bibitem [{\citenamefont {Jolly}\ and\ \citenamefont
		{Porras}(2024)}]{jolly24-1}%
	\BibitemOpen
	\bibfield  {author} {\bibinfo {author} {\bibfnamefont {S.~W.}\ \bibnamefont
			{Jolly}}\ and\ \bibinfo {author} {\bibfnamefont {M.~A.}\ \bibnamefont
			{Porras}},\ }\href@noop {} {\bibfield  {journal} {\bibinfo  {journal}
			{Journal of the Optical Society of America B}\ }\textbf {\bibinfo {volume}
			{41}},\ \bibinfo {pages} {577} (\bibinfo {year} {2024})}\BibitemShut
	{NoStop}%
	\bibitem [{\citenamefont {Porras}\ and\ \citenamefont
		{Jolly}(2025)}]{porras25-2}%
	\BibitemOpen
	\bibfield  {author} {\bibinfo {author} {\bibfnamefont {M.~A.}\ \bibnamefont
			{Porras}}\ and\ \bibinfo {author} {\bibfnamefont {S.~W.}\ \bibnamefont
			{Jolly}},\ }\href@noop {} {\bibfield  {journal} {\bibinfo  {journal}
			{Physical Review A}\ }\textbf {\bibinfo {volume} {053523}},\ \bibinfo {pages}
		{111} (\bibinfo {year} {2025})}\BibitemShut {NoStop}%
	\bibitem [{\citenamefont {Bor}\ \emph {et~al.}(1993)\citenamefont {Bor},
		\citenamefont {R{\'a}cz}, \citenamefont {Szab{\'o}}, \citenamefont
		{Hilbert},\ and\ \citenamefont {Hazim}}]{bor93}%
	\BibitemOpen
	\bibfield  {author} {\bibinfo {author} {\bibfnamefont {Z.}~\bibnamefont
			{Bor}}, \bibinfo {author} {\bibfnamefont {B.}~\bibnamefont {R{\'a}cz}},
		\bibinfo {author} {\bibfnamefont {G.}~\bibnamefont {Szab{\'o}}}, \bibinfo
		{author} {\bibfnamefont {M.}~\bibnamefont {Hilbert}},\ and\ \bibinfo {author}
		{\bibfnamefont {H.~A.}\ \bibnamefont {Hazim}},\ }\href@noop {} {\bibfield
		{journal} {\bibinfo  {journal} {Optical Engineering}\ }\textbf {\bibinfo
			{volume} {32}},\ \bibinfo {pages} {2501} (\bibinfo {year}
		{1993})}\BibitemShut {NoStop}%
	\bibitem [{\citenamefont {Pretzler}\ \emph {et~al.}(2000)\citenamefont
		{Pretzler}, \citenamefont {Kasper},\ and\ \citenamefont
		{White}}]{pretzler00}%
	\BibitemOpen
	\bibfield  {author} {\bibinfo {author} {\bibfnamefont {G.}~\bibnamefont
			{Pretzler}}, \bibinfo {author} {\bibfnamefont {A.}~\bibnamefont {Kasper}},\
		and\ \bibinfo {author} {\bibfnamefont {K.~J.}\ \bibnamefont {White}},\
	}\href@noop {} {\bibfield  {journal} {\bibinfo  {journal} {Applied Physics
				B}\ }\textbf {\bibinfo {volume} {70}} (\bibinfo {year} {2000})}\BibitemShut
	{NoStop}%
	\bibitem [{\citenamefont {Torres}\ \emph {et~al.}(2010)\citenamefont {Torres},
		\citenamefont {Hendrych},\ and\ \citenamefont {Valencia}}]{torres10}%
	\BibitemOpen
	\bibfield  {author} {\bibinfo {author} {\bibfnamefont {J.~P.}\ \bibnamefont
			{Torres}}, \bibinfo {author} {\bibfnamefont {M.}~\bibnamefont {Hendrych}},\
		and\ \bibinfo {author} {\bibfnamefont {A.}~\bibnamefont {Valencia}},\ }\href
	{https://doi.org/10.1364/AOP.2.000319} {\bibfield  {journal} {\bibinfo
			{journal} {Advances in Optics and Photonics}\ }\textbf {\bibinfo {volume}
			{2}},\ \bibinfo {pages} {319} (\bibinfo {year} {2010})}\BibitemShut {NoStop}%
	\bibitem [{\citenamefont {Hyde}\ and\ \citenamefont {Porras}(2023)}]{hyde23}%
	\BibitemOpen
	\bibfield  {author} {\bibinfo {author} {\bibfnamefont {M.~W.}\ \bibnamefont
			{Hyde}}\ and\ \bibinfo {author} {\bibfnamefont {M.~A.}\ \bibnamefont
			{Porras}},\ }\href@noop {} {\bibfield  {journal} {\bibinfo  {journal}
			{Physical Review A}\ }\textbf {\bibinfo {volume} {108}},\ \bibinfo {pages}
		{013519} (\bibinfo {year} {2023})}\BibitemShut {NoStop}%
	\bibitem [{\citenamefont {Novikov}(2025)}]{novikov25}%
	\BibitemOpen
	\bibfield  {author} {\bibinfo {author} {\bibfnamefont {V.~B.}\ \bibnamefont
			{Novikov}},\ }\href@noop {} {\bibfield  {journal} {\bibinfo  {journal}
			{Optics Letters}\ }\textbf {\bibinfo {volume} {50}},\ \bibinfo {pages} {1540}
		(\bibinfo {year} {2025})}\BibitemShut {NoStop}%
	\bibitem [{\citenamefont {Gao}\ \emph {et~al.}(2025)\citenamefont {Gao},
		\citenamefont {Chen}, \citenamefont {Guo}, \citenamefont {Fan}, \citenamefont
		{Chen}, \citenamefont {Lu},\ and\ \citenamefont {Hu}}]{gaoX25}%
	\BibitemOpen
	\bibfield  {author} {\bibinfo {author} {\bibfnamefont {X.}~\bibnamefont
			{Gao}}, \bibinfo {author} {\bibfnamefont {W.}~\bibnamefont {Chen}}, \bibinfo
		{author} {\bibfnamefont {Y.}~\bibnamefont {Guo}}, \bibinfo {author}
		{\bibfnamefont {J.}~\bibnamefont {Fan}}, \bibinfo {author} {\bibfnamefont
			{W.}~\bibnamefont {Chen}}, \bibinfo {author} {\bibfnamefont {Y.}~\bibnamefont
			{Lu}},\ and\ \bibinfo {author} {\bibfnamefont {M.}~\bibnamefont {Hu}},\
	}\href {https://doi.org/10.1117/1.APN.4.3.036003} {\bibfield  {journal}
		{\bibinfo  {journal} {Advanced Photonics Nexus}\ }\textbf {\bibinfo {volume}
			{4}},\ \bibinfo {pages} {036003} (\bibinfo {year} {2025})}\BibitemShut
	{NoStop}%
\end{thebibliography}
\end{document}